%% file: Q-in-crowdsourcing.tex
\documentclass[prodmode,acmcsur]{acmsmall} 
\usepackage{longtable}
\usepackage[ruled]{algorithm2e}
\usepackage{color}
\usepackage{caption}
\usepackage{wrapfig}
\usepackage{multirow}
\usepackage{array}

\newcommand*\rot{\rotatebox{90}}
\newcolumntype{L}[1]{>{\raggedright\let\newline\\\arraybackslash\hspace{0pt}}m{#1}}

\SetAlFnt{\small}
\SetAlCapFnt{\small}
\SetAlCapNameFnt{\small}
\SetAlCapHSkip{0pt}
\IncMargin{-\parindent}

\newcommand{\rev}[1]{{#1}}

\acmVolume{51}
\acmNumber{1}
\acmArticle{7}
\acmYear{2018}
\acmMonth{01}

\begin{document}

\markboth{F. Daniel et al.}{Quality Control in Crowdsourcing: A Survey}

\title{Quality Control in Crowdsourcing: A Survey of Quality Attributes, Assessment Techniques and Assurance Actions}

\author{Florian Daniel
\affil{Politecnico di Milano}
Pavel Kucherbaev
\affil{Delft University of Technology}
Cinzia Cappiello
\affil{Politecnico di Milano}
Boualem Benatallah
\affil{University of New South Wales}
Mohammad Allahbakhsh
\affil{University of Zabol}}

\begin{abstract}
Crowdsourcing enables one to leverage on the intelligence and wisdom of potentially
large groups of individuals toward solving problems. Common problems
approached with crowdsourcing are labeling images, translating or transcribing text,
providing opinions or ideas, and similar -- all tasks that computers are not good at or
where they may even fail altogether. The introduction of humans into computations
and/or everyday work, however, also poses critical, novel challenges in terms of quality 
control, as the crowd is typically composed of people with unknown and 
very diverse abilities, skills, interests, personal objectives and technological resources. 
This survey studies quality in the context of crowdsourcing along several dimensions, 
so as to define and characterize it and to understand the current state of the art. 
Specifically, this survey derives a quality model for crowdsourcing tasks, identifies 
the methods and techniques that can be used to assess the attributes of the model,
and the actions and strategies that help prevent and mitigate quality problems. 
An analysis of how these features are supported by the state of the art further
identifies open issues and informs an outlook on hot future research directions.
\end{abstract}




\acmformat{F. Daniel, P. Kucherbaev, C. Cappiello, B. Benatallah, M. Allahbakhsh.
Quality Control in Crowdsourcing: A Survey of Quality Attributes, Assessment 
Techniques and Assurance Actions.}

\begin{bottomstuff}
The work of B. Benatallah is supported by the ARC (Australian Research Council) Discovery Project Grant No. DP1601104515: Integrating Quality Control into Crowd-Sourcing Services.

Author's addresses: F. Daniel and C. Cappiello are with the Politecnico di Milano, DEIB, Via Ponzio 34/5, 20133 Milano, Italy; emails: \{florian.daniel, cinzia.cappiello\}@polimi.it; P. Kucherbaev is with EEMCS, Web Information System, P.O. Box 5031, 2600 GA Delft, The Netherlands; email: p.kucherbaev@tudelft.nl; B. Benatallah is with the University of New South Wales, CSE, Sydney, NSW 2052, Australia; email: b.benatallah@unsw.edu.au; M. Allahbakhsh is with the University of Zabol, Computer Department, Faculty of Engineering, Zabol, Iran; email: allahbakhsh@uoz.ac.ir.
\end{bottomstuff}

\maketitle

\input{sections/1.intro}

\input{sections/3.quality}

\input{sections/4.framework}
\input{sections/5.model}

\input{sections/6.assessment}

\input{sections/7.assurance}

\input{sections/8.soa}

\input{sections/10.discussion}

\bibliographystyle{ACM-Reference-Format-Journals}
\bibliography{qcsurvey,survey}

\elecappendix

\input{sections/A.references}

\end{document}

%% file: sections/1.intro.tex

\section{Introduction}

\emph{Crowdsourcing} is the outsourcing of a piece of work
to a crowd of people via an open call for contributions \cite{csfirst}. 
In crowdsourcing, one group of people (so-called \emph{requesters}) 
submit \emph{tasks} to a crowdsourcing \emph{platform} (or service); 
another group of people (the \emph{workers} that form the \emph{crowd}) 
contribute to solving the task. The result of solving the task is called an 
\emph{output}. Requesters may evaluate outputs and \emph{reward} 
workers depending on the respective quality; in 
situations where requesters delegate responsibility for 
quality control to the crowdsourcing platform, outputs may be checked and 
rewarded directly and automatically by the crowdsourcing service itself. 
\emph{Rewards} can be money, gifts, reputation badges or similar \cite{minder2012crowdlang}.

Depending on the task to crowdsource and the acceptance criteria by 
both the requester and the workers to enter a mutual business relationship, 
different negotiation models 
have emerged so far: The \emph{marketplace model} \cite{Ipeirotis2010b} 
targets so-called micro-tasks of limited complexity, such as tagging a picture 
or translating a piece of text, for which the requester typically requires a large 
number of answers. Prominent examples of crowdsourcing platforms that 
implement the marketplace model are Amazon Mechanical Turk, 
Microworkers and CrowdFlower. The \emph{contest model}
\cite{CavalloAAMAS2012} is particularly suitable to creative works
where the requester fixes the budget he is willing to spend and workers
compete with their solutions for the reward. Examples are 
99designs, InnoCentive and IdeaScale. The \emph{auction model} 
\cite{Satzger2013} targets works where the requester fixes the acceptance 
criteria and workers bid for the task. An example of auction platform is 
Freelancer. But also \emph{volunteering}, e.g., like in Wikipedia or 
Crowdcrafting, has proven its viability, and the spectrum of variations of 
these models is growing.

The critical aspect is that outputs produced by the crowd must be checked for \emph{quality}, 
since they are produced by workers with unknown or varied skills and motivations 
\cite{minder2012crowdlang,malone2010collective}. The quality of a crowdsourced 
task is multifaceted and depends on the quality of the workers involved in its
execution, the quality of the processes that govern the creation of tasks, the 
selection of workers, the coordination of sub-tasks like reviewing intermediary outputs, 
aggregating individual contributions, etc.
A large body of empirical studies confirms that existing crowdsourcing 
platforms are not robust to effectively check and control the quality of 
crowdsourced tasks or to defend against attacks such as cheating, 
manipulating task outputs, or extracting sensitive information from 
crowdsourcing systems \cite{kritikos2013survey}. Concerns about the 
unintended consequences of poor quality control methods, including 
financial, intellectual property and privacy risks, malicious attacks, and 
project failure are growing \cite{minder2012crowdlang,malone2010collective,kritikos2013survey}.
 
\rev{To be fair, it is important to note that platform
providers may not have the necessary information about tasks to approach
these concerns appropriately. For instance, in marketplace platforms task
design is typically fully under the control of the requester, and the platform
is not aware if the tasks raises intellectual property issues or not. 
Some quality control and submission filtering activities may therefore also
be carried out by the requester outside the platform to meet expected
levels of quality. Task-specific platforms, such as 99designs for graphical 
design tasks, instead, are well aware of the problem (e.g., intellectual property 
rights) and help requesters and workers to manage them. They can do so, 
as they focus on few specific task types only, which they know and can 
support well.}

The increasing importance of crowdsourcing services and the intensification 
of global competition indicate however that building proper solutions
to quality problems should now be a top priority. Ideally, developers of 
crowdsourcing applications should be offered effective quality control 
techniques, methods and tools that let them systematically build, verify 
and change quality control mechanisms to suit their requirements. Yet,
no framework exists that endows crowdsourcing services with robust and 
flexible quality control mechanisms, and most research on quality control 
in crowdsourcing has focused on single, specific aspects of quality only, 
such as worker reputation or redundancy. In addition, conceived quality 
control techniques are typically embedded inside proprietary platforms and 
not generalizable. Consequently, designing, building and maintaining robust 
and flexible crowdsourcing quality controls remains a deeply challenging 
and partly unsolved problem, and requesters and platform operators 
may not have the knowledge, skills or understanding to craft an own 
quality-aware strategy to safely take advantage of the opportunities offered 
by crowdsourcing. For instance, \citeN{stol2014two} describe a case study 
on crowdsourced software development that shows how unpreparedness to
handle quality issues quickly increased project costs. \citeN{lasecki2014information} 
show how state-of-the-art crowd-powered systems are still not ready to deal with 
``active attacks'', while \citeN{gadiraju2015understanding} show that malicious 
workers can easily cause harm if suitable quality controls are missing.

This survey aims to shed light on the problem of quality control in crowdsourcing
and to help users of crowdsourcing services and developers of crowdsourcing
applications to understand the various moments where quality comes into
play in the crowdsourcing process, how it is manifest (or not), and how it
can be assessed and assured via suitable methods and actions. 
Concretely, it provides the following contributions:

\begin{itemize}
\item An \emph{introduction} to quality control in crowdsourcing (Section 
\ref{sec:quality}) and a \emph{taxonomy} to understand and classify 
the state of the art in quality control techniques for crowdsourcing. 
The taxonomy focuses on the three core aspects of quality:

\begin{itemize}
\item The \emph{quality model} that emerges from the state of the art 
(Section \ref{sec:model}), that is, the dimensions and attributes to describe 
quality in crowdsourcing services.
\item The \emph{assessment} (the measuring and its methods) of the values of the 
attributes identified by the quality model (Section \ref{sec:assessment}).
In order to instantiate the quality model, it is necessary to know and
master the respective assessment methods.
\item The \emph{assurance} of quality (Section \ref{sec:assurance}), that is,
the set of actions that aim to achieve expected levels of 
quality. To prevent low quality, it is paramount to understand how to design
for quality and how to intervene if quality drops below expectations. 
\end{itemize}

\item A comprehensive \emph{analysis} of how state-of-the-art crowdsourcing 
platforms and services support quality control in practice (Sections \ref{sec:analysis}). 

\item A discussion of shortcomings and limitations along with the
respective \emph{challenges and opportunities} for future research and development 
(Section \ref{sec:discussion}).
\end{itemize}

%% file: sections/3.quality.tex
\section{Quality in crowdsourcing}
\label{sec:quality}

\begin{figure}[t]
\centering
\includegraphics[scale=0.6]{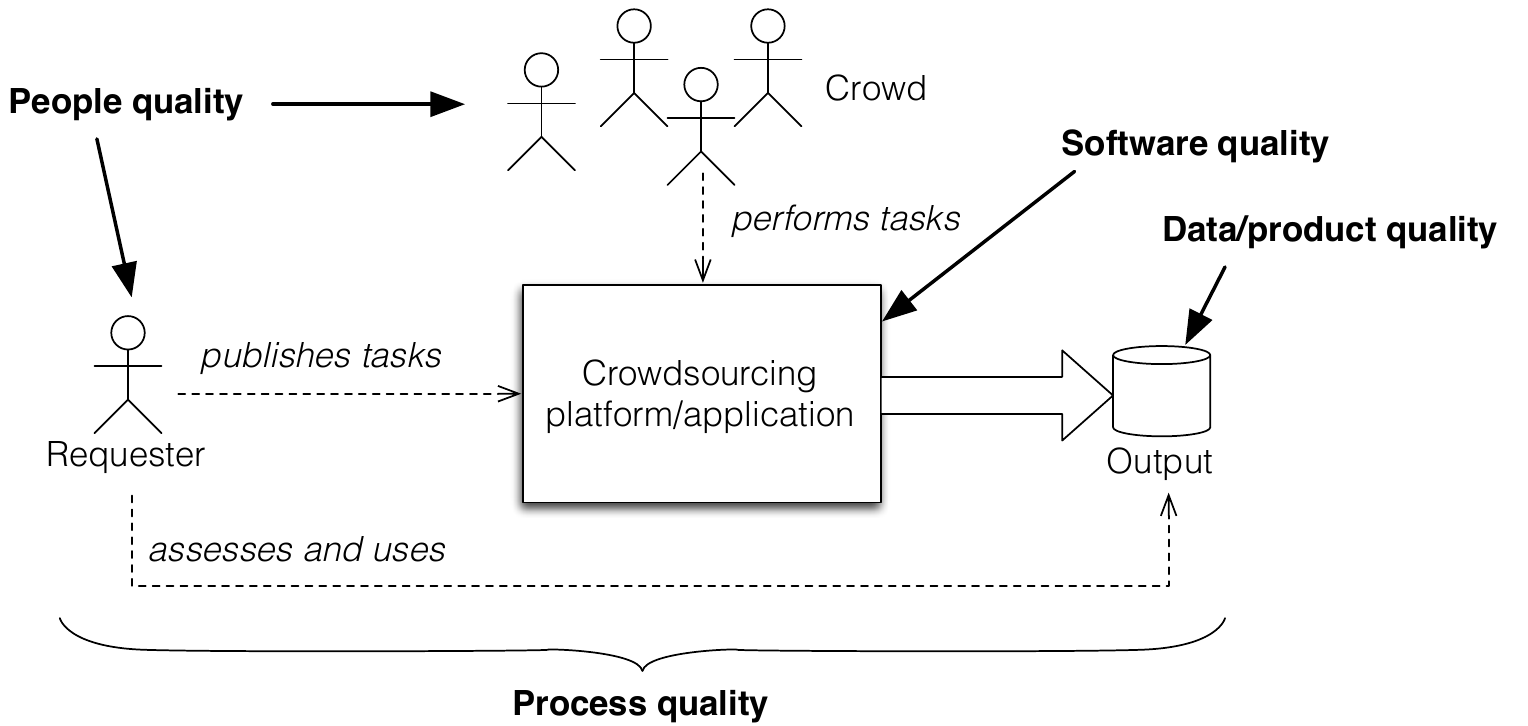}
\caption{The basic crowdsourcing scenario with its quality aspects.}
\label{fig:scenario}
\end{figure}

Quality (and its control) is an active research area in several computing disciplines, including data quality \cite{batini2009methodologies,Hunter2013}, quality of software products \cite{herbsleb1997software,jung2004measuring}, 
software services \cite{kritikos2013survey}, and user interfaces \cite{hartmann2008towards}. 
In many cases, research from these fields is complementary to quality control in crowdsourcing, and some elements can be adopted to leverage application-specific quality control mechanisms (e.g., some code quality metrics can be leveraged in programming crowdsourcing tasks).
However, the peculiarities of crowdsourcing require taking into account novel quality issues that are rising, especially 
to consider the user involvement in the execution of different tasks, e.g., production of contents, annotations, and evaluations. 

Therefore, it is necessary to consider that in crowdsourcing systems the quality of the output is influenced by multiple factors (see Figure \ref{fig:scenario}): The requester defines and publishes tasks to be executed by the crowd via a crowdsourcing platform or a dedicated application. The crowd produces outputs that are checked for quality and delivered to the requester. 
The output quality depends, among others, on the profiles and abilities of workers, the description of the tasks, the incentives provided, the processes implemented to detect malicious behaviors or low-quality data, as well as on the reputation of and collaboration by the requester.
The quality of the output can therefore be expressed by means of quality dimensions that model objective and subjective aspects of \emph{data and product quality}, while the actual quality of the outputs is influenced by aspects related to \emph{people} (the workers, requesters and possible other actors), \emph{software} (the crowdsourcing platform or application and the design of tasks) and \emph{process quality} (the organization of work and the implemented quality measures).

Adapting the expectation-centric definition of quality (e.g., \cite{LewisBooms83}) to crowdsourcing, we define the quality of a crowdsourced task as \textit{the extent to which the output meets and/or exceeds the requester's expectations.}

%% file: sections/4.framework.tex
\subsection{Taxonomy}
\label{sec:framework}


Considering the current fragmented literature and lack of an 
all-encompassing view of quality control in crowdsourcing, 
we developed the taxonomy depicted in Figure 2 to understand 
and analyze how quality is dealt with in crowdsourcing.  The 
proposed taxonomy features a holistic view on the problem of 
quality control in crowdsourcing and aims to highlight the key 
aspects and options one has to face when developing quality
control mechanisms. The taxonomy is based on our own 
experience 
\cite{Allahbakhsh2013,CappielloDKMP11,kritikos2013survey,batini2009methodologies}
as well as on an
extensive literature review of related areas, discussions with 
colleagues, experimentation with systems and 
prototypes, which allowed us to identify common building 
blocks for the different variations in quality control mechanisms. 
Accordingly, we propose a taxonomy with three 
categories that may, in turn, be split into sub-categories:

\begin{figure}[t]
\centering
\includegraphics[scale=0.6]{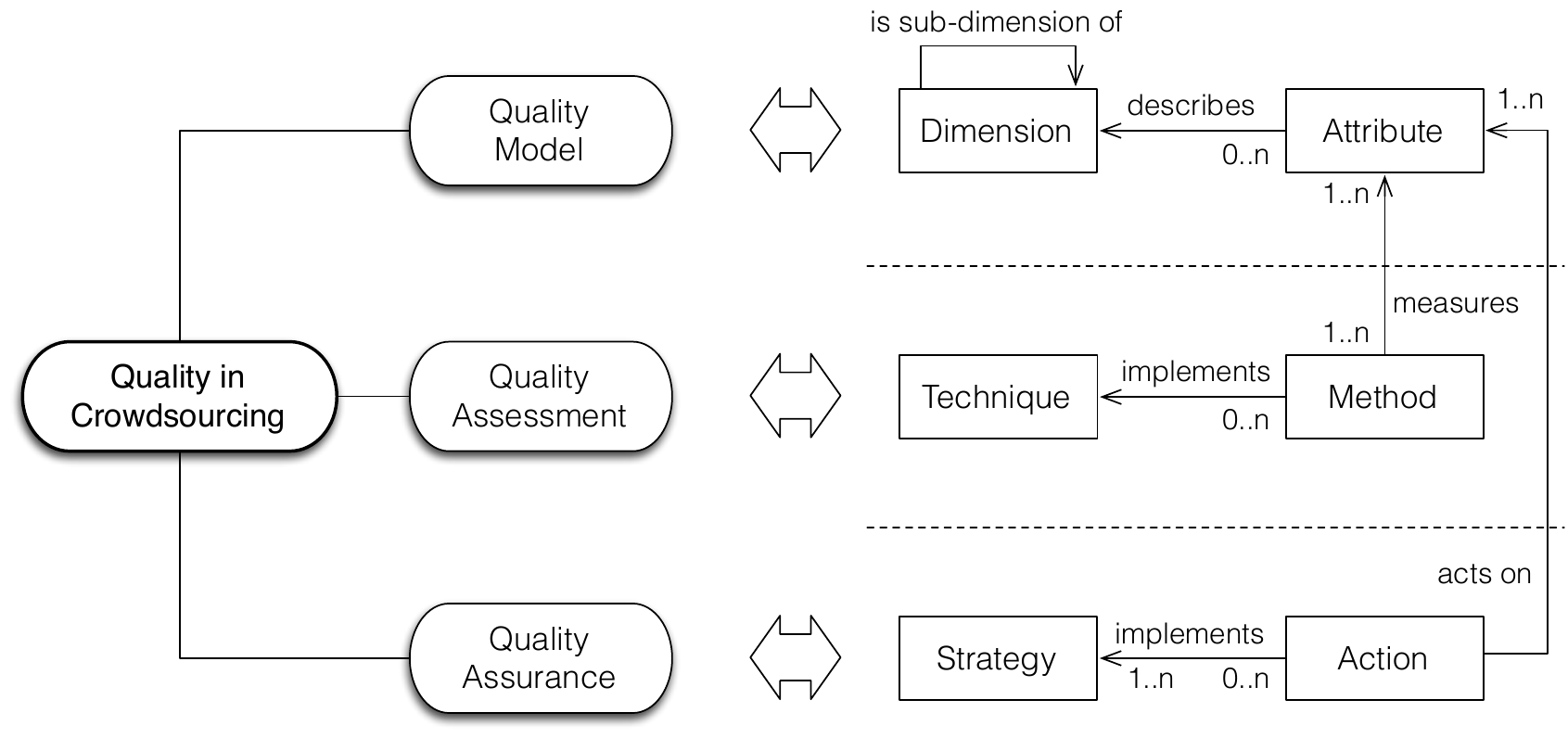}
\caption{The components of the quality control taxonomy 
and their internal structures.}
\label{fig:framework}
\end{figure}

\begin{itemize}

\item A \emph{quality model} for crowdsourcing that captures which 
quality dimensions and attributes have been identified so far in the literature.

\begin{itemize}
\item \emph{Dimensions} represent the components that constitute 
a crowdsourcing task, such as the input and output data of the task 
or the people involved in the task. Dimensions are described by attributes 
and are not directly measurable.
\item \emph{Attributes} characterize properties (qualities) of 
the task, such as the accuracy of data or the expertise of workers.
Attributes are \emph{concrete} if they are measurable; they are 
\emph{abstract} if they are not directly measurable and values are 
derived from concrete attributes, e.g., aggregations.
\end{itemize}

\item An analysis of the \emph{quality assessment}  
methods that have been used so far to assess quality attributes
in the context of crowdsourcing.
 
\begin{itemize}
\item \emph{Techniques} distinguish who performs the assessment
from a high level of abstraction. For instance, techniques that involve single
individuals (e.g., rating) differ from techniques that involve groups of people
(e.g., voting).
\item \emph{Assessment methods} allow one to measure quality 
attributes. For example, accuracy may be measured comparing
outputs with a ground truth, while worker expertise may be measured 
through questionnaires. Methods are like functions that are executed 
automatically, manually or both and produce a value as output.
\end{itemize}

\item A study of the \emph{quality assurance} actions that 
allow one to improve quality by acting on the quality attributes
identified in the quality model.

\begin{itemize}
\item \emph{Strategies} represent the top-level decisions to be 
taken when aiming at improving quality, 
that is, what to act upon and in which direction. For instance, selecting 
good workers and training workers are two different strategies that aim 
to improve the quality of the people involved in a task.
\item \emph{Actions} are the basic operations one can perform to
prevent or fix quality problems, such as 
checking credentials or showing a training video. Each action
implements a specific strategy.
\end{itemize}

\end{itemize}

In the following sections, we detail each of these components and
explain them with the help from the respective literature.


%

\subsection{Literature Selection}

In order to identify the references to consider in this survey, we selected
a set of conferences and journals that, to the best of our knowledge,
publish research on crowdsourcing and related topics. The conferences 
considered were: \rev{AAAI, BPM, CAiSE, CHI, CI, CIKM, CSCW, ECML, ECSCW, HCOMP, ICML, ICSE, ICWE, iUI, KDD, NIPS, SIGIR, UBICOMP, UIST, VLDB, WSDM, and WWW}. The journals considered were: \rev{ACM CSUR, ACM TIIS, ACM TOCHI, ACM TOIS, ACM TOIT, ACM TOSEM, ACM TWEB, Communications of the ACM, CSCW, Information Systems, IEEE Computer, IEEE Internet Computing, IEEE TKDE, IEEE TSC, IEEE TSE, VLDB, and WWW}. In order to keep the 
selection of references manageable and up to date, we queried for 
contributions from 2009 onwards using the following keywords
in either title, abstract or keywords: 
Crowd, Crowdsourcing, Human Computation, Collective Intelligence, Social Computing,
Collaborative Computing, Collaborative Systems, 
Wikinomics, Mass Collaboration, Micro-tasking,
Crowd Labour. Articles were retrieved through the advanced 
search feature of the ACM Digital Library and SCOPUS. Papers published 
in HCOMP and Collective Intelligence were retrieved manually, as at
the time of querying they were not properly indexed by any digital 
library. The selection specifically looked for conference and journal 
papers, and neglected demo papers, posters and workshop papers. 
The search identified \rev{1013} papers. A further manual
check filtered out \rev{257} papers that we finally considered relevant for this 
survey. Additional papers considered stem from prior knowledge by the authors. 


%
%
%

%% file: sections/5.model.tex

\section{Crowdsourcing Quality Model}
\label{sec:model}

Figure \ref{fig:qualitymodel} illustrates the quality model identified as a result of 
this survey. We identify the following dimensions to group the quality attributes 
in crowdsourcing systems:

\begin{itemize}

\item \emph{Data}: This refers to the data required to perform a task or 
produced as a result of performing a task by a worker, i.e, task input and
output data. For instance, input data can be 
images to label or a text to translate, and the corresponding output data 
can be the labels of the images and the translated text. Quality control in 
crowdsourcing all revolves around achieving high-quality output data, which 
is the core challenge for mass adoption of crowd work \cite{kittur2013future}.

\item \emph{Task}: 
The type of work proposed, the way it is described and how it is implemented
strongly affects the attractiveness of a task and the quality of the outputs that 
can be achieved. We distinguish the following sub-dimensions:

\begin{itemize}
\item \emph{Description}: This is the description of the work the requester 
asks the crowd to perform; it includes the instructions of 
how to perform the task and possible context information. 
The clarity and details of the description influence the way workers 
perform the task and hence the quality of its output \cite{chandler2013risks}.

\item \emph{User interface}: This is the software user interface workers use to
perform the task. In most tasks, this interface is a simple HTML form through 
which workers submit their contributions. But it can also come in the form of
a stand-alone application with input techniques not supported by standard 
HTML forms (e.g., selecting patterns in a model). The quality of the user interface
determines how easily workers can perform tasks \cite{marcus2012counting}.

\item \emph{Incentives}: Crowdsourcing generally implies paid, online work.
The incentives, that is, the stimuli the requester offers to attract workers to
tasks, plays therefore a crucial role in crowdsourcing. Incentives may come
in two different forms: extrinsic incentives (e.g., rewards) and intrinsic incentives
(e.g., worker status). Several studies have identified direct relations between
incentives and output quality and/or task execution speed \cite{singer2013pricing}.

\item \emph{Terms and conditions}: These are the general rules and arrangements
that govern the work relationship between the requester and the workers. Aspects
like the privacy of workers, the protection of intellectual property (IP), compliance 
with laws, ethical standards, and similar are typically specified here. Terms 
and conditions affect worker interest and legal aspects \cite{wolfson2011look}. 

\begin{figure}[t]
\centering
\includegraphics[scale=0.6]{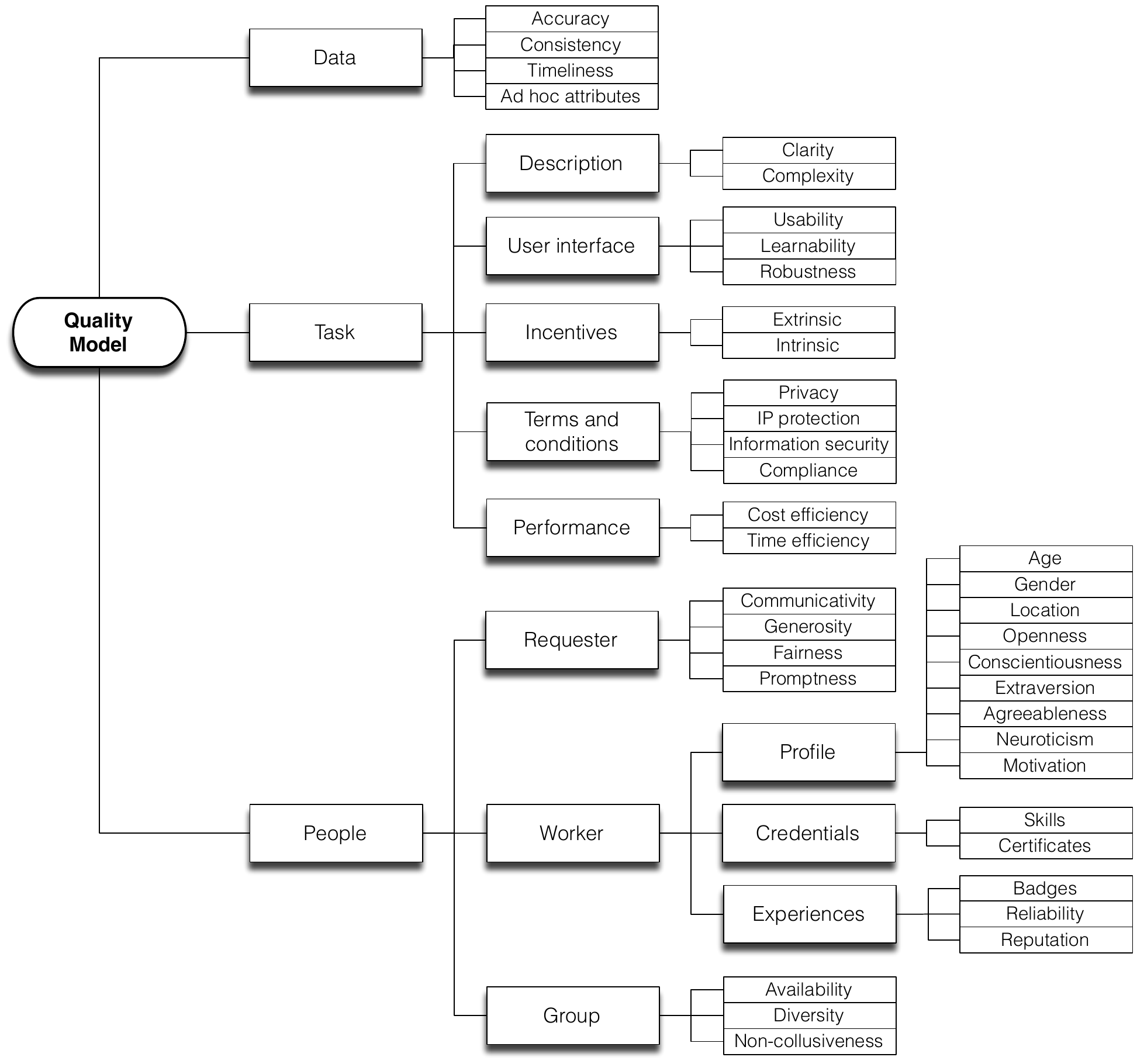}
\caption{The quality model for crowdsourcing tasks emerged from literature
with \emph{dimensions} (boxes with shadow) and \emph{attributes} 
(boxes without shadow).}
\label{fig:qualitymodel}
\end{figure}

\item \emph{Performance}: Performance expresses the
amount of useful work completed in the context of a crowdsourced
task compared to the time and resources needed for its processing.
Crowdsourcing can be an intricate endeavor, with large volumes of input 
data to be processed and hundreds or thousands of workers involved,
that can easily grow complex. Controlling the consumption
of resources is of utmost importance to guarantee the sustainability of
crowdsourcing initiatives.
\end{itemize}

\item \emph{People}: These are the actors involved in the task, including requesters, 
workers, and the crowd. Good work requires qualified, 
prepared and trusted people. In some tasks, workers may also be required to collaborate
with other workers or requesters. We distinguish 
three sub-dimensions regarding the people involved in crowdsourcing:

\begin{itemize}
\item \emph{Requester}: The first sub-dimension acknowledges the role of the
requester, who is not only the one that crowdsources work, but also the one
that may evaluate outputs and interact with workers. Fairness and communication are examples of attributes of a good requester \cite{irani2013turkopticon,trustcol12}.

\item \emph{Worker}: The other natural sub-dimension focuses on the workers.
Substantial work has been done on discriminating good from bad workers. We
group the identified attributes into profile, credentials, and experiences, 
in function of what the attributes describe. \emph{Profile}  characterizes an
individual's identity (age, gender, location) and describes the individual's 
distinctive character, typically expressed through \rev{motivation and} so-called
personality traits, i.e., relatively stable, enduring properties of people that can
influence behavior \cite{john2008paradigm}. \emph{Credentials} are documented
qualifications that people bring into a system from the outside. 
\emph{Experiences} express knowledge acquired or social relations
established by using a system; they are usually
based on system-specific metrics and rules.

\item \emph{Group}: Finally, people can be studied also in groups, e.g., the
crowd as a whole or smaller teams of collaborating workers. Good groups,
for instance, are formed by non-colluding workers \cite{KhudaBukhsh2014} 
or by workers that are able to perform also very diverse tasks with good quality.

\end{itemize}

\end{itemize}

We describe the attributes that characterize each dimension next. 
Appendix \ref{app:references} summarizes the referenced literature.

\subsection{Data Quality}
A large 
body of work focused on characterizing and handling data quality in the
context of crowdsourcing. The most important attribute studied is the
\emph{accuracy} of output data \cite{hansen2013quality,Kazai2011,Yin2014}. Several synonyms are used to refer to
accuracy, such as ``goodness'' \cite{Cao2014}, ``correctness'' \cite{Yin2014}, 
or just generically ``quality'' \cite{Eickhoff2012,Kulkarni2012,Kern2010}.
Data \emph{consistency} is commonly interpreted as the similarity between 
outputs produced by different workers in response to a same input. It has been 
studied, for instance, in the context of peer consistency evaluation
\cite{huang2013enhancing,Eickhoff2012}. The \emph{timeliness} of data 
is the property of outputs to be available in useful time for further processing 
\cite{kittur2013future}. Timeliness is especially important in near-realtime
crowdsourcing scenarios \cite{lasecki2013real,lasecki2013warping} and 
has also been studied under the name of ``reaction time'' \cite{Yin2014}.

It is important to note that next to these quality attributes that characterize 
the quality of a dataset in general terms, 
many crowdsourcing scenarios ask for the use of \emph{ad-hoc attributes} 
that are able to capture task-specific properties. For instance, 
\citeN{yu2014comparison} study the accuracy, coverage, and conciseness 
of textual summaries; the latter two attributes are specific to the problem 
of summarizing texts. \citeN{nguyen2014using} distinguish between
13 different features to assess narrative similarity. 

\subsection{Task Description Quality}
Regarding the quality of task descriptions, its \emph{clarity}
is of utmost importance \cite{Hossfeld2014}. \citeN{Tokarchuk2012} 
state that clarity positively correlates with performance, a result that 
is empirically confirmed by \citeN{georgescu2012map}, while 
\citeN{kulkarni2012b} specifically study the problem of incomplete 
descriptions. Needless to say, if a worker does not immediately understand
a task, he will not be willing to work on it. 
Several authors have studied the \emph{complexity} of 
tasks (or ``granularity'' \cite{hu2012deploying}), identified correlations 
with worker motivation \cite{rogstadius2011assessment}, 
and matched workers with tasks based on complexity scores
\cite{difallah2013pick}.

\subsection{User Interface Quality}
A user-friendly, understandable interface can attract more workers and 
increase the quality of outputs \cite{Allahbakhsh2013}. In this respect,
especially \emph{usability} as been studied, e.g., for workers from low-income 
countries \cite{khanna2010evaluating}, in photographing tasks
\cite{noronha2011platemate}, or in task design \cite{retelny2014expert}. 
\citeN{alagarai2014cognitively} specifically focus on visual saliency and
working memory requirements as sub-properties of usability.
\citeN{willett2012strategies} show that the design of the task
interface has an impact on the \emph{learnability} of a task. A good 
task interface is further characterized by a high \emph{robustness}
against cheaters, that is, it is able to produce high-quality outputs even
in the presence of cheaters \cite{Eickhoff2012}. \citeN{Hung2013} 
talk about ``sensitivity to spammers".

\subsection{Incentives}
\emph{Incentives} affect the attractiveness of a task. As \citeN{Hossfeld2014} 
point out, there are several possible incentives targeting either the extrinsic
(reward-driven) or intrinsic (interest-driven) motivation of workers. According
to the authors, increasing extrinsic motivation leads to faster task completion,
while increasing intrinsic motivation leads to higher quality. Many researchers
have specifically studied the role of monetary rewarding in crowdsourcing, 
e.g., on speed \cite{heer2010crowdsourcing} or execution efficiency 
\cite{singer2013pricing}, while \citeN{Eickhoff2012} compared fun and 
social prestige with monetary rewards as incentives.

\subsection{Terms and Conditions}
At a more abstract level, \emph{privacy} has been identified as 
key property of tasks that deal with personal data, e.g., using images 
that show people \cite{lasecki2013real} \rev{or asking workers to share
their position \cite{Boutsis2016}}. But also \emph{information
security} and \emph{IP protection}, that is, the protection of data and
intellectual property (IP), are emerging quality attributes that affect a requester's
willingness to crowdsource \cite{towardCS}. A requester may, for example, 
share source code or design documents, which are assets that contain IP; 
quality control mechanisms are needed, e.g., to limit 
the access of workers to information, invite vetted workers only, or 
sign nondisclosure agreements. 
\rev{As \citeN{Amor2016} point out in their approach to crowdsourced team 
competitions, the problem is not limited to requesters only and may affect also workers.}
More generically, \emph{compliance}
means conformance with laws and regulations \cite{wolfson2011look}, but also
with commonly accepted user policies \cite{wang2012serf} (e.g., no malicious 
crowdsourcing campaigns) or expected ethical behaviors by the requester 
\cite{irani2013turkopticon}. If a task is perceived as non-compliant by workers,
it is unlikely they will perform it.

\subsection{Task Performance}
The two most important attributes that have been studied to capture the
execution performance of a task are cost efficiency and time efficiency.  
The expected cost of a task is easily determined by multiplying the reward by
the number of task instances worked on \cite{ambati2012,Livshits2014}.
However, since crowdsourcing a task is a generally non-deterministic process,
\emph{cost efficiency}, i.e., the cost per completed task instance or the cost 
per correct output, has been studied more intensively
\cite{ipeirotis2014quizz,rokicki2014competitive}. The \emph{time efficiency}
can be defined as the number of tasks completed in a given temporal interval
\cite{Eickhoff2012}; \citeN{lin2014signals} use the synonym ``throughput,''
\citeN{hung2013evaluation} talk about ``computation time'' for a given set of
tasks. \citeN{KucherbaevCSCW2016}, for instance, aim to improve the time 
efficiency by re-launching tasks at runtime. \rev{\citeN{Cheng2015a} compute 
task effort based on error rates and task completion times.}

\subsection{Requester Reputation}
\citeN{irani2013turkopticon} propose Turkopticon, a browser extension 
for Firefox and Chrome that augments workers' view of their task list in
Amazon Mechanical Turk with information other workers have provided 
about requesters. Turkopticon supports assessing requesters by means
of four attributes: 
\emph{Communicativity} captures how responsive a requester is to 
communications or concerns raised by a worker.
\emph{Generosity} tells how well a requester pays for the amount of time 
necessary to complete a task.
\emph{Fairness} (also studied by \citeN{trustcol12}) tells how fair a requester is in approving or rejecting work 
submitted by workers.
\emph{Promptness} captures how promptly a requester approves and pays work 
that has been successfully submitted.

\subsection{Worker Profile}
\citeN{kazai2011worker} study different worker profile attributes and personality 
traits in labeling tasks. \emph{Age}, for instance, correlates with output accuracy,
while \emph{gender} does not seem to have any influence on quality. \emph{Location}
does impact quality \cite{kazai2011worker,Eickhoff2012,kazai2012face}.
To assess the personality of workers, \citeN{kazai2011worker} use the so-called 
Big Five inventory and study five personality traits (definitions by
\citeN{john2008paradigm} and correlations from \cite{kazai2011worker}):
\emph{Openness} ``describes the breadth, depth, originality, and complexity of an
individual's mental and experiential life;'' it correlates with output accuracy.
\emph{Conscientiousness} ``describes socially prescribed impulse control that
facilitates task- and goal-directed behavior, such as thinking before acting;''
it positively correlates with output accuracy.
\emph{Extraversion} ``implies an energetic approach toward the social and material 
world and includes personality traits such as sociability, activity, assertiveness, 
and positive emotionality;'' it negatively correlates with output accuracy.
\emph{Agreeableness} ``contrasts a prosocial and communal orientation toward 
others with antagonism and includes traits such as altruism, tender-mindedness,
trust, and modesty;'' it correlates with output accuracy.
\emph{Neuroticism} ``contrasts emotional stability and even-temperedness with
negative emotionality, such as feeling anxious, nervous, sad, and tense;''
it negatively correlates with output accuracy. According to \citeN{kazai2012face},
personality characteristics are useful to distinguish between good and better 
workers (less between good and bad). \rev{\citeN{Kobayashi2015} introduce a taxonomy of worker \emph{motivations}, so that appropriate incentives could be applied to persuade workers to contribute}.

\subsection{Worker Credentials}
Credentials are all those qualifications or personal qualities that 
describe a worker's background; they can be self-declared (e.g., programming
language skills) or issued by official bodies (e.g., a MSc degree 
is issued by a university). \emph{Skills} are abilities that tell if 
a worker is able to perform a given task. \rev{\citeN{Mavridis2016} propose a taxonomy of skills}; \citeN{difallah2013pick} 
use skills to match workers and tasks; \citeN{schall2014crowdsourcing}
identify skills automatically. \emph{Certificates} are documents
that attest skills, such as academic certificates or language certificates
\cite{Allahbakhsh2013}.

\subsection{Worker Experience}
\emph{Badges} are platform-provided certificates
of performance, e.g., performing a certain number of actions or tasks
of a given type \cite{anderson2013steering}. Badges can be used 
to select workers, but also to motivate them: badges are seen as
``virtual trophies'' \cite{scekic2013incentives}.
The \emph{reliability} of a worker (often also called ``accuracy'' of the worker) 
is commonly interpreted as the aggregated accuracy of the outputs produced 
by the worker \cite{Kazai2011}, or the worker's error rate in answering questions
\cite{dalvi2013aggregating,demartini2013large}, or the acceptance rate of 
outputs delivered by workers (e.g., Mechanical Turk). \citeN{Sakurai2013} use the
synonym ``correctness'' of a worker as the probability of the worker being
correct in labeling tasks. \rev{\citeN{Raykar2011} assign scores to workers based on how reliable (not random) their answers are}. The \emph{reputation} of a worker may take into 
account additional parameters, such as the worker's timeliness, the quality of 
the evaluators that assessed the worker, relations with other workers or 
requesters, the trust they express toward the worker, and similar
\cite{Allahbakhsh2013}. That is, reputation also captures other 
community members' feedback about a worker's activity in the system \cite{de2011reputation}.

\subsection{Group Quality}
The quality of groups of people, i.e., teams or the crowd as a whole, has
been studied mostly in terms of three different aspects. \emph{Availability},
i.e., the presence in a platform of enough workers or experts with the 
necessary skills for a given task, has been identified as an issue \cite{li2014wisdom}.
Low availability usually leads to low quality of outputs \cite{ambati2012} 
or slow task execution \cite{li2014wisdom}.
Next, \emph{diversity} is the property of a group to represent different
types of people, skills, opinions, and similar. Diversity is particularly important
if representative samples of people are searched for, e.g., in tasks that ask
for opinions like the ox weight estimation experiment by \citeN{crowdwisdombook}
or polls \cite{Livshits2014}. \citeN{willett2012strategies} specifically 
study how to increase the diversity of worker outputs.
Finally, \emph{non-collusiveness} means that a group of agents 
does not share information or make joint 
decisions contrary to explicit or implicit social rules, which would 
result in an unfair advantage over non-colluding agents or other 
interested parties \cite{KhudaBukhsh2014}.

%% file: sections/6.assessment.tex
\section{Quality Assessment}
\label{sec:assessment}

\begin{figure}[t]
\centering
\includegraphics[scale=0.6]{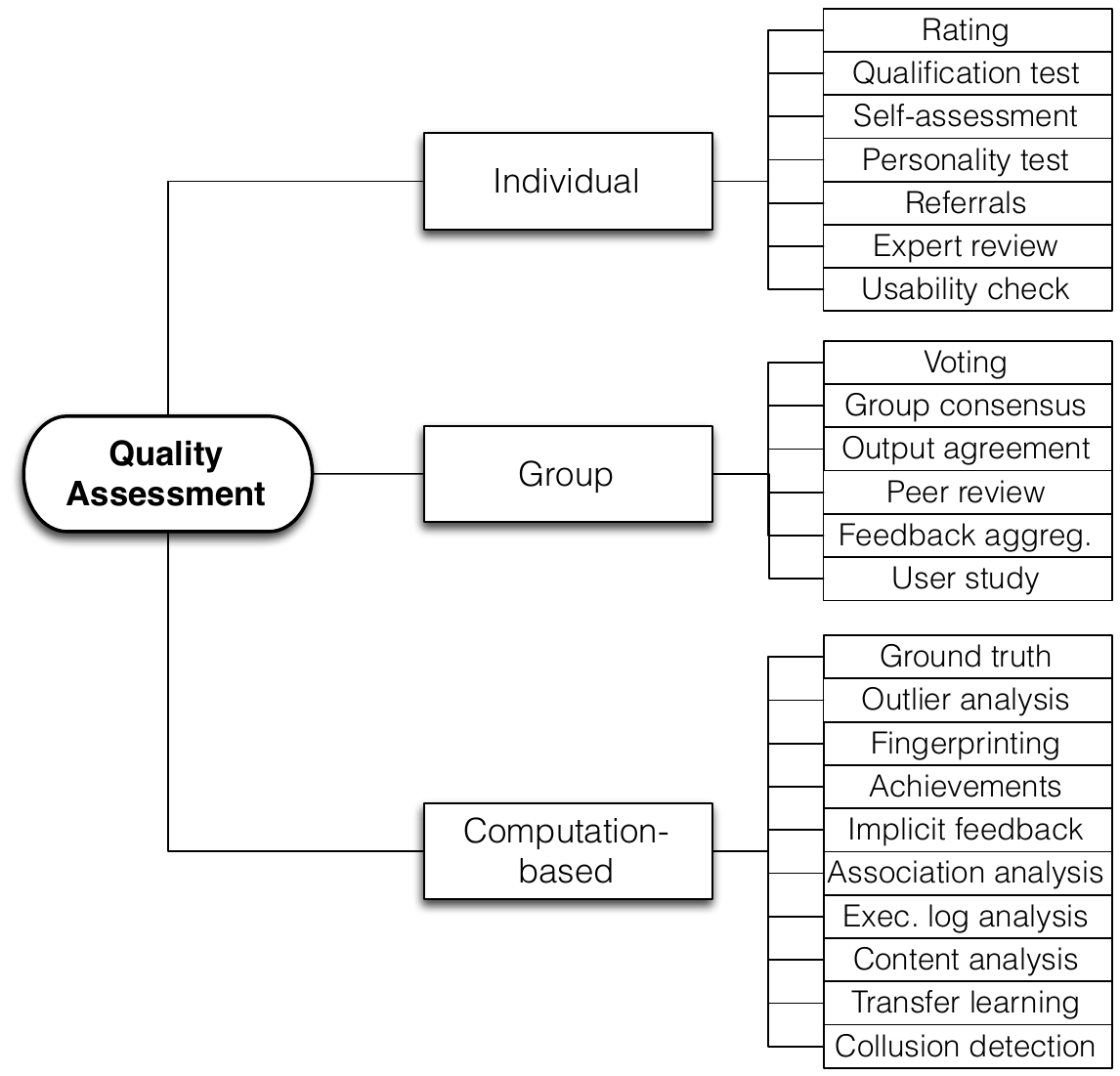}
\caption{The quality assessment \emph{methods} (on the right) 
emerged from literature; the boxes with shadow represent generic \emph{techniques}.}
\label{fig:assessmodel}
\end{figure}

Assigning concrete values to some of the attributes identified above can be 
a straightforward exercise, e.g., verifying if a worker has the necessary
skills, certificates or badges for a task can easily be done manually or 
automatically \cite{Khazankin2012}, or a task duration can easily be read from the 
log of the crowdsourcing system. Assessing some other attributes may instead
require the use of methods based on both automated metrics as well 
as manual interventions by workers, the requester and/or external experts.
In the following, we focus on these more complex methods, and organize 
the discussion of the identified assessment methods into three major groups 
(the techniques) in function of the actor in the assessment task, as illustrated
in Figure~\ref{fig:assessmodel}:

\begin{itemize}

\item \emph{Individual}: Given the human-oriented nature
of crowdsourcing, it is natural to think about involving humans also
into assessment tasks and not only into work tasks. 
Some assessment methods require the involvement of individuals (workers, experts,
or the requester), such as rating the accuracy of a given output
or writing a review.

\item \emph{Group}: Some other assessment methods require the joint 
action of a group of people (typically workers) for the formation of an
assessment. For instance, voting requires multiple participants to 
derive a ranking from which to select the best, or peer review 
requires multiple peers to judge the work of a colleague.

\item \emph{Computation-based}: Some other assessment methods,
instead, can be performed without the involvement of humans, that
is, automatically by a machine. Comparing a set of outputs with a
given ground truth can, for example, be carried out automatically if suitable comparison operators are available. 

\end{itemize} 

We describe the respective methods next, following the order
proposed in Figure~\ref{fig:assessmodel}. We summarize the 
discussion tabularly in Appendix \ref{app:references}.

%
%
%
%

\subsection{Individual Assessments}

\subsubsection{Rating}
Rating means assigning a value to an item chosen from a scale 
to express the perceived quality of the item. The scale
defines the possible values to choose from, e.g., unary scales
allow one to express if an item is liked or not (used for instance
in \rev{Instagram}), binary scales distinguish between two values 
(good/bad, true/false, thumbs-up/thumbs-down, or similar), 
\rev{ordinal} scales distinguish between discrete sets of positive 
and negative values (typically 5, 7, or more), and continuous scales may 
allow an arbitrary number of values inside an interval ([0..1], 
[1..100], or similar). Rating is extensively used in crowdsourcing
for quality assessment \cite{dalvi2013aggregating,yu2014comparison},
text similarity and understandability \cite{nguyen2014using}, worker 
confidence \cite{Sakurai2013}, task quality (in CrowdFlower workers rate tasks
after completion), and requester reputation \cite{irani2013turkopticon}. 
\rev{\citeN{Hata2017} have demonstrated that it is possible to predict workers' 
long-term quality using just a glimpse of their quality on the first five tasks
as rated by the requester.}
\rev{Instead, to counteract reputation inflation (too
generous ratings, even if the rater would not want to work with the rated
worker/requester again), \citeN{Gaikwad2016} conceived a rating
system that rebounds the consequences of feedback back onto the
raters, e.g., by giving positively rated workers precedence in future tasks
by the same requester.}

\subsubsection{Qualification test}
A qualification test is a questionnaire that a worker may be required
to fill to obtain access to tasks that require prior knowledge or skills, 
e.g., language skills or acquaintance with a programming or modeling 
tool. Thanks to the a-priori knowledge of the correct answers of the 
test, qualification tests can be evaluated automatically, grating the 
worker access to the task if a minimum threshold of correct answers is
reached. Qualification tests are widespread in crowdsourcing and many
crowdsourcing platforms provide support for qualification tests.
\citeN{heer2010crowdsourcing} studied the effectiveness of qualification
tests to assess workers in graphical perception experiments;
their findings show that qualification tests are indeed able to discriminate
unprepared workers, thereby increasing output quality.

\subsubsection{Self-assessment}
Self-assessment asks workers to assess the quality of their own work
after producing an output for a task. The practice stems from 
self-assessment in learning \cite{boud2013enhancing}, where the aim is
to help students reflect, learn, and connect their work better with learning
goals. \citeN{Dow2012} studied the effectiveness of self-assessment in 
the context of crowdsourced product reviews
and found that tasks with self-assessment produced better overall
quality than tasks without. In addition, they found that self-assessment
helped workers improve the quality of their work over time, in line
with the expected learning effect. Self-assessment has also been used
to allow workers to state the confidence they have in their own work
\cite{Sakurai2013}\rev{, even suggesting workers to skip the task if they are not confident \cite{Shah2015}}. 
\rev{\citeN{ShahZ16} show noisy output examples to workers right before task submission so they can resolve possible ``silly'' mistakes.} 

\subsubsection{Personality tests}
These are tests, typically questionnaires shown to workers, that
allow one to assess personality traits, that is, the actions, attitudes and
behaviors individuals possess. For instance, \citeN{kazai2011worker} and \citeN{kazai2012face} used the so-called Big Five inventory \cite{john1999big} to assess the personality 
traits openness, conscientiousness, agreeableness, extraversion, and 
neuroticism; the authors further identified a positive relation between
the former three and output accuracy and a negative relation between
the latter two and output accuracy. Turkopticon proposed by 
\cite{irani2013turkopticon}, instead, allows workers to explicitly rate
the communicativity, generosity, fairness, and promptness of requesters.

\subsubsection{Referrals}
Referrals express that someone has referred someone else for consideration. 
They are well known in the domain of recruitment as a 
way to gather and confirm expertise. SalesForce has indicated in their official blog that their main strategy for recruitment is based on referrals
(\url{https://www.salesforce.com/blog/2015/01/behind-scenes-salesforce-our-1-recruiting-secret.html}). 
LinkedIn and ResearchGate allow tagging skills to professionals.
While referrals may not be a common instrument to identify workers, 
they may well be used to find experts. Facebook and Twitter 
enable recruiting workers by exploiting social connections \cite{bozzon2012answering}.

\subsubsection{Expert review}
An expert review is an assessment provided by a person that is considered
a domain expert by the requester. This expert is commonly not a
member of the crowd (whose work the expert assesses) and is typically
directly assigned by the requester to assessment tasks. \citeN{Dow2012} 
describe a system that supports expert reviews providing workers with 
feedback on the quality of their work and show that expert reviews help 
increase quality. \rev{Expert reviews are however costly; accordingly, 
\citeN{Hung2015} devised a method that optimizes the effort of experts 
through reviews of only partial worker output sets while keeping quality high.}

\subsubsection{Usability check}
Checking the usability of a task design helps identify issues that may
prevent workers from producing high-quality outputs. Specifically, 
usability guidelines \cite{nielsen2002homepage} can be used by the 
requester to check if a task design follows known best practices or not.
In the specific context of crowdsourcing, \citeN{willett2012strategies} 
have studied the usability of task UIs (understandability and learnability) 
for data analysis and identified seven guidelines for data analysis tasks:
use feature-oriented prompts, provide good examples, include reference 
gathering subtasks, include chart reading subtasks, include annotation 
subtasks, use pre-annotated charts, and elicit explanations iteratively.

\subsection{Group Assessments}

\subsubsection{Voting}
Voting means expressing preference for one or more candidates (e.g., outputs)
out of a group of candidates; aggregating the votes by multiple voters enables
the derivation of a list of candidates ranked by preference. Voting is 
very common in crowdsourcing and used to make group decisions, 
e.g., 2 out of 3 majority decisions. \citeN{Kulkarni2012}, for instance, 
provide built-in support for voting in their collaborative crowdsourcing
platform Turkomatic to validate the quality of task outputs.  
\citeN{little2010turkit} instead equip their human programming framework
Turkit with a dedicated \emph{vote} programming construct.
\citeN{caragiannis2014modal} study different voting 
techniques for crowdsourcing, while \citeN{sun2012majority} also point 
out some pitfalls of voting that should be taken into account when using 
the technique.

\subsubsection{Group consensus}
Group consensus is similar to voting, yet, the consensus refers to ratings 
assigned to an item and less to a mere expression of preference. The purpose
is therefore not ranking multiple items (outputs) but identifying the most 
representative rating for one item. The method is frequently used to produce 
consensus labels from crowd data. \citeN{Sheshadri2013} compare different
techniques to assess offline consensus and study, given multiple noisy 
labels (or ratings) per item, how to infer the best consensus label.
\rev{\citeN{ZhangTKDE2015} specifically address the problem of imbalanced
labeling, i.e., labeling scenarios where there are many more positive than 
negative cases (or viceversa).}
\citeN{Eickhoff2012} use disagreement with the majority consensus to 
identify workers considered cheaters (if they disagree more then 67\%
of the times with the majority) and to derive a measure of robustness of
a task. \rev{However, consensus must not always be good: \citeN{Kairam2016},
for instance, show that there is also value in divergent outputs in crowdsourcing.}

\subsubsection{Output agreement}
An output agreement is reached if two or more workers, given the same
input and performing the same task, produce a same or similar result as 
output. \citeN{Waggoner2014}, for instance, study the use of output
agreement to assess worker reliability and show that the technique is particularly 
useful to elicit common knowledge (since the workers know that they
are assessed based on the similarity of their outputs with those by other
workers). \citeN{huang2013enhancing} use output agreement 
(they use the term peer consistency) to assess and motivate workers. 
\rev{\citeN{Jagabathula2014} instead assess workers based on output disagreement.}

\subsubsection{Peer review}
Peer review is similar to expert reviews, with the key difference that it
involves multiple peers in the assessment, in order to limit the
bias of individual peers and to elicit an as correct as possible assessment.
It is typically used in those situations where experts would not be able
to assess alone all outputs or items, such as in the paper selection of
scientific conferences, and represents a form of community-based
self-assessment. Crowdsourcing shares similar characteristics.
\citeN{hansen2013quality}, for example, use peer review in the context 
of the FamilySearch Indexing project and show that peer review is more
efficient than using an arbitrator to resolve disagreements among 
volunteers. \citeN{zhu2014reviewing} study different peer reviewing
approaches and show that the practice, next to representing an effective 
assessment instrument, may also lead to performance increases of the 
reviewers (the workers themselves). 
\rev{Peer review among workers has also been 
successfully used as reputation assessment technique, producing
better results than conventional techniques \cite{Whiting2017}.}

\subsubsection{Feedback aggregation}
More sophisticated aggregation algorithms can be used to integrate
large amounts of feedbacks provided by either workers or requesters, 
in order to obtain representative, concise assessment of quality attributes.
\citeN{dalvi2013aggregating}, for instance, propose an eigenvector-based 
technique to estimate both worker reliabilities and output qualities.
Simple analytic models (e.g., sum, average, minimum or maximum) are 
employed to calculate the rating scores of products. In weighted averaging 
techniques, evaluations cast by users are weighted and their impact on final 
rating scores is adjusted based on their corresponding weights. 
\rev{\citeN{Davtyan2015} leverage on content similarity to increase the 
quality of aggregated votes compared to standard majority voting approaches.}
\citeN{trustcol12} use time and credit of crowdsourcing tasks to weight 
pairwise evaluations between the community members and also propose
the concept of fairness of an evaluator while evaluating other members.
Iterative approaches calculate weights of evaluations and the result of 
aggregation simultaneously but in several iterations. Iterative methods 
for social rating have been pioneered in \cite{iterativefilter1,iterativefilter2}.
\citeN{AleksRep} have proposed a reputation calculation model for online 
markets. Many more examples of aggregation algorithms exist 
\cite{hung2013evaluation}, with \citeN{joglekar2013evaluating} also
generating confidence intervals for aggregated values. 

\subsubsection{User study}
Assessing the effectiveness of task UIs for which there do not yet exist
reference guidelines may require conducting task-specific usability studies. 
\citeN{willett2012strategies}, for instance, conducted user studies directly
on Amazon Mechanical Turk without direct contact between the experimenters
and the workers. 
\citeN{khanna2010evaluating}, instead, organized controlled between-subjects 
study sessions with workers, implemented dedicated image annotation tasks,
and observed them in action to assess the usability of tasks to low-income 
workers from countries like India. The key general barriers identified were 
the complexity of instructions and UIs, difficult navigation, and different 
cultural backgrounds. \citeN{alagarai2014cognitively} used eye tracking to
measure the working memory required by different task designs and their
visual saliency. This kind of user studies requires expert skills, for example, for the
conduct of interviews, the design of questionnaires or proper user observations.

\subsection{Computation-based Assessments}

\subsubsection{Ground truth}
The use of ground truth data (gold data, control questions) is a common approach 
in crowdsourcing: by injecting into tasks questions whose answers are known 
and formalized a priori (so that they can be checked automatically), it is 
possible to computationally estimate the aggregate accuracy of outputs and the
trust in workers. Ground truth evaluation is considered one of the most 
objective mechanisms that can accurately measure the performance of workers
\cite{huang2013enhancing}. The method is, for instance, natively supported 
by CrowdFlower. \citeN{Eickhoff2012} provide an example of how to use ground
truth data in gamification. \citeN{hara2013combining} provide good examples of 
how to collect ground truth answers for image labeling with wheelchair 
drivers (domain experts), while \citeN{oleson2011programmatic} argue that 
the generation of ground truth data is very difficult and costly 
and propose the use of ``programmatic gold'' generated from previously 
collected correct data. \rev{\citeN{Liu2013} propose a method predicting an optimal number of ground truth labels to include}. \citeN{le2010ensuring} study the distribution logics 
of gold questions and conclude that a uniform distribution produces best results 
in terms of worker precision, \rev{while \citeN{ElMaarry2015} show that the biggest
threat to ground truth evaluations are tasks with highly skewed answer distributions}. 
CAPTCHAs can be used to tell human workers
and machines (e.g., robots) apart \cite{lasecki2014information,von2008recaptcha}.

\subsubsection{Outlier analysis}
Outliers are data points (e.g., worker performance or estimations of a property)
that significantly differ from the remaining data \cite{aggarwal2013introduction},
up to the point where they raise suspicion. Outliers may thus identify poorly
performing workers, random answers, or similar. In \cite{rzeszotarski2012crowdscape},
the authors show how they use CrowdScape to link behavioral worker information
with output properties and identify outliers visually (in charts). \citeN{jung2011improving}
use outlier analysis to identify ``noisy workers'' in consensus labeling tasks.
\citeN{marcus2012counting} use outlier analysis to detect ``spammers'' in 
estimation tasks.

\subsubsection{Fingerprinting}
This method captures behavioral traces from workers during task execution
and uses them to predict quality, errors, and the likelihood of cheating. 
Behavioral traces are identified by logging user interactions with the task UI 
(at the client side) and are expressed as interaction patterns that
can be used at runtime to monitor a worker's conformance with the patterns.
The method has been coined by \citeN{rzeszotarski2011instrumenting},
demonstrating the effectiveness of the approach in predicting output quality.
\rev{\citeN{Kazai2016} even conclude that ``accuracy almost doubles in some 
tasks with the use of gold behavior data.''} As an extension,
in \cite{rzeszotarski2012crowdscape} the authors propose CrowdScape, a 
system ``that supports the human evaluation of complex crowd work through 
interactive visualization and mixed initiative machine learning.''

\subsubsection{Achievements}
Verifying the achievement of predefined goals is a typical method used to 
assign so-called badges (certificates, proves of achievement) to users.
The method resembles the merit badge concept of the Scout movement,
where one must attain all preceding badges before qualifying for the next 
one \cite{6125717}. Achievements are used in crowdsourcing 
especially in the context of gamified crowdsourcing tasks, such as exemplarily
shown in \cite{Massung2013}, where badges are used in a mobile data collection
application to engage casual participants in pro-environmental data collection. 
Badges were earned for activities such as using the application for five days 
in a row or for rating a shop on the weekend.

\subsubsection{Implicit feedback}
Implicit feedback is a method of content-based feedback analysis. Feedback 
is implicit and extracted from the behavior of evaluators, rather than from 
explicit feedback forms \cite{de2011reputation}. For example, in WikiTrust 
\cite{adler2007content,adler2011wikipedia}, a reputation management tool 
designed for assessing Wikipedia users, the reputation of the user is built based 
on the changes a user makes to the content. If a change is 
preserved by editors, the user gains reputation otherwise he 
looses reputation. \citeN{lin2014signals} analyze implicit signals about task 
preferences (e.g., the types of tasks that have been available and displayed 
and the number of tasks workers have selected and completed) to recommend
tasks to workers. \citeN{difallah2013pick} analyze workers' personal interests
expressed in social networking sites (Facebook) to recommend tasks of likely
interest in Mechanical Turk.

\subsubsection{Association analysis}
Associations among people are, for instance, the friend-of relationships in 
Facebook, the recommendations in Freelancer, or the following relationship
in Github. These relationships can be interpreted as expressions of trust or
prestige and analyzed accordingly. Already in 1977, \cite{freeman1977set} 
proposed the idea of betweenness centrality as a measure of a node'��s importance 
inside a graph, with a special attention toward communicative potential. 
In the specific context of crowdsourcing, \citeN{RajasekharanMN13} have, for
example, extended the well-known Page Rank algorithm \cite{page1999pagerank}
with edge weights to compute what the authors call a ``community activity rank'' 
for workers to eventually assess their reputation. This kind of network analysis
techniques is typically more suited to contest crowdsourcing models, in which
workers may know each other, and less to marketplaces where there is no 
communication among workers.

\subsubsection{Task execution log analysis}
Given a log (trace) of worker interactions and task completions, it is possible
to analyze and/or mine the log for assessment purposes. Fingerprinting uses
log analysis to identify patterns; here the purpose is measuring quality attributes. 
\citeN{KucherbaevCSCW2016}, for example, use linear regression
to estimate task duration times at runtime to identify likely abandoned tasks, 
i.e., tasks that will not be completed by their workers. \rev{\citeN{Moshfeghi2016}
use game theory to classify workers based on task execution durations}. 
Going beyond the 
estimation of individual quality attributes, \citeN{heymann2011turkalytics} 
propose Turkalytics, a full-fledged analytics platform for human computation
able to provide real-time insight into task properties like demographics of
workers as well as location- and interaction-specific properties.
\citeN{Huynh2013} reconstruct from logs a provenance network that captures
which worker saw/produced which data item in a set of interrelated tasks;
the network allows the authors to predict the trustworthiness of task outputs.
\citeN{Jung2014} predict output quality by analyzing the time series of workers'
past performance; \citeN{KhudaBukhsh2014} do so to identify colluding workers.

\subsubsection{Content analysis}
It is also possible to automatically analyze a task's description and text 
labels to assess properties like task difficulty or trust in the requester. 
\citeN{artz2007survey}, for instance, speak about ``common sense'' 
rules to make trust decisions, e.g., do no trust prices below 50\% of the 
average price. \citeN{difallah2013pick} propose the use of different methods
to assess task difficulty: comparison of task description with worker skills,
entity extraction and comparison (based on Linked Open Data), or content-oriented
machine learning algorithms.
\rev{Also \cite{yang2016modeling} come to the conclusion that ``appearance 
and the language used in task description can accurately predict task complexity.''}
 \citeN{alagarai2014cognitively}, instead, analyzed
the semantic similarity of input field labels and showed that too diverse labels
may act as distractors and that these can be used to predict accuracy.

\subsubsection{Transfer learning}
Transfer learning is the improvement of learning in a new task through the 
transfer of knowledge from a related task that has already been learned 
\cite{torrey2009transfer}. In crowdsourcing, transfer learning has been 
used to borrow knowledge from auxiliary historical tasks to improve the 
data veracity in a target task, for instance, to infer the gender or reliability 
of workers using a hierarchical Bayesian model \cite{mo2013cross}. 
\citeN{fang2014active} use transfer learning to estimate labelers' expertise
in data labeling tasks inside Mechanical Turk. \citeN{zhao2013transfer},
instead, apply transfer learning in a cross-platform setting to transfer
knowledge about workers from Yahoo! Answers to a Twitter-based 
crowdsourcing system.

\subsubsection{Collusion detection}
Collusion detection aims to identify groups of colluding workers, i.e.,
workers that share information to gain an advantage. \citeN{KhudaBukhsh2014}
aim to identify non-adversarial collusions in ratings by detecting 
strong inter-rater dependencies across tasks, as these diverge from 
the mean ratings. \citeN{marcus2012counting} inject ground truth data to
detect colluders. \citeN{Allahbakhsh2014SOCA}, instead, compute a probability
of collusion by analyzing past collaborations of workers on same tasks and 
recommend likely non-colluding workers for team formation.

%% file: sections/7.assurance.tex

\section{Quality Assurance}
\label{sec:assurance}

The logical step after assessing quality is assuring quality,
that is, putting into place measures that help achieve quality --
the more so if the assessment unveiled low quality for any of the
attributes identified earlier. These measures, concretely, come in 
the form of strategies and actions that a requester may implement.
Some of these strategies and actions are \emph{reactive} if they react 
to clearly identified quality issues, e.g., filtering outputs upon the 
verification that some outputs do not meet given quality thresholds.
Other strategies and actions are instead \emph{proactive} in that
they do not need a triggering event to be applied, e.g., following
proper usability guidelines in the implementation of a task does 
not require a prior verification of usability.

Before looking into the strategies and actions that have been 
used so far for quality assurance in crowdsourcing, it is important
to note that already assessing (measuring) quality, especially if 
the object of the assessment are people, may have positive side
effects on quality. Most notably, reviewing has been shown to 
impact positively the performance of both workers and reviewers
\cite{zhu2014reviewing} and quality in general \cite{hansen2013quality}.
Rating has been used to increase the requesters awareness of 
workers concerns and rights \cite{irani2013turkopticon}, but also
rating the performance of workers has similar positive
side effects \cite{Dow2012}. Many other studies that provide 
similar evidence exist. 

In the following, we do not further study these side effects of
assessment. Instead, we review the strategies and actions that
specifically aim to improve quality as first-order goal. Many of
them require a prior quality assessment (e.g., filtering outputs
requires first assigning quality labels to them), others do not. We
explain these aspects next, and organize the discussion as illustrated
in Figure \ref{fig:assurance}. The identified \emph{strategies} are:

\begin{figure}[t]
\centering
\includegraphics[scale=0.6]{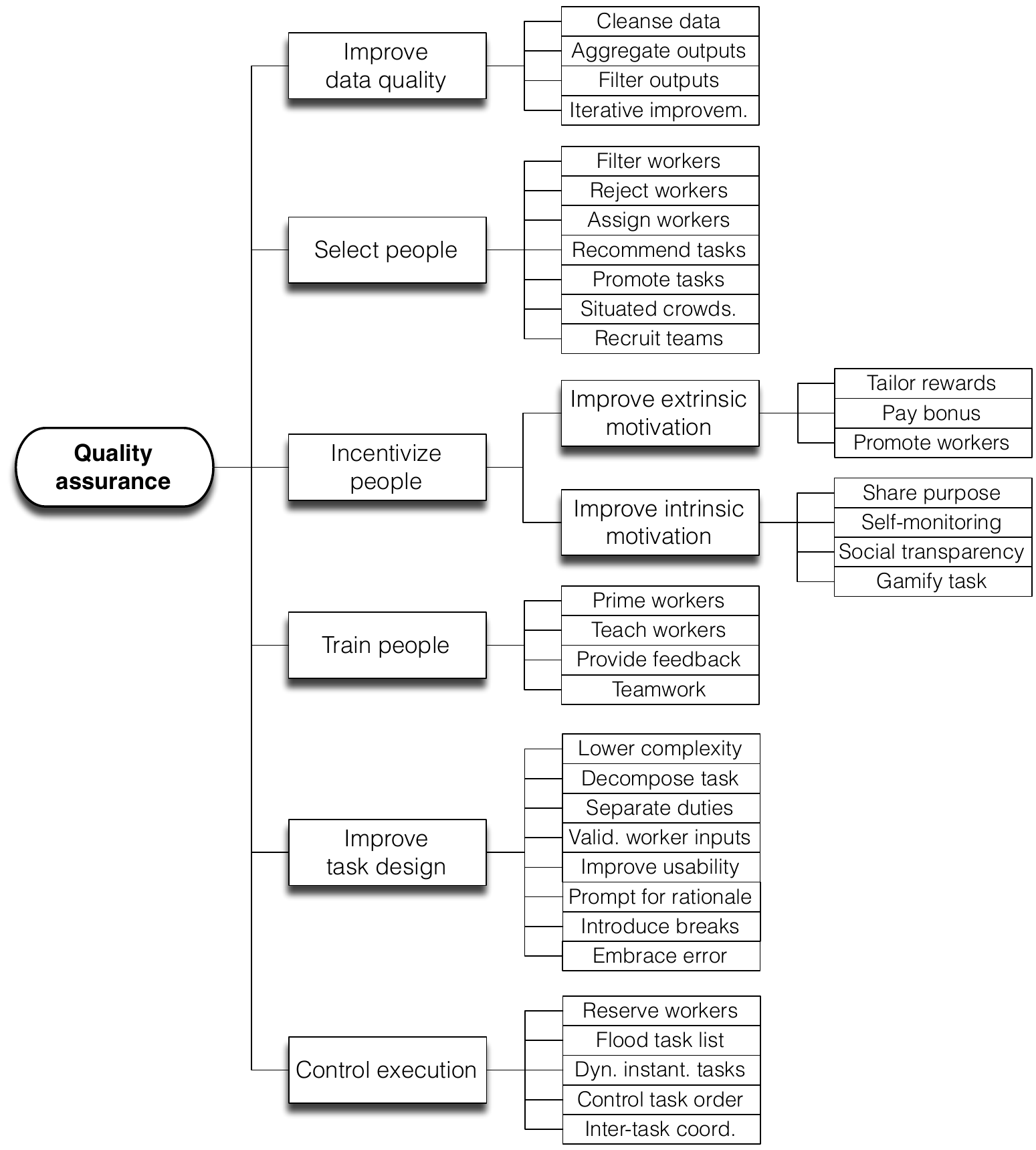}
\caption{The quality assurance \emph{strategies} (boxes with shadow) 
and \emph{actions} (boxes without shadow) emerged from literature.}
\label{fig:assurance}
\end{figure}

\begin{itemize}

\item \emph{Improve data quality}: The first and most intuitive
strategy to approach low quality in the output data is improving 
the quality of the data itself, where possible. Typical actions 
range from cleansing inputs (crowdsourcing is not immune to the
garbage-in/garbage-out problem) to the iterative improvement of
outputs.

\item \emph{Select people}: Another intuitive strategy is to identify
workers that produce better results. Doing so may require requesters
to filter workers by their profiles, recommending tasks to workers 
they think they will perform well or eliminating cheaters.

\item \emph{Incentivize people}: Incentivizing people means acting
on the motivation that pushes people to perform well (or not). There
are two different sub-strategies that aim to leverage on two different 
types of drivers \cite{rogstadius2011assessment}:

\begin{itemize}

\item \emph{Extrinsic motivation} depends on drivers that
are determined from the outside of a person's own control. For example,
workers may work harder if they see that it results into a higher 
reward or even a bonus from the requester.

\item \emph{Intrinsic motivation}, instead, depends on drivers
that are internal to the person and does not depend on the desire for
an external reward. For instance, workers may be pushed to perform
better if they can compare their performance with that of other workers
\rev{or if performing a task is entertaining (like in games)}.

\end{itemize}

\item \emph{Train people}: Workers can also be instructed or trained
to be prepared better for specific tasks or domains and, hence,
to perform better. Different approaches can be followed, such
as teaching workers or providing feedback to the
work they submit.

\item \emph{Improve task design}: One reason for low quality may be
the low usability or understandability of the user interface workers are
provided with or the task description and structure
itself. Improving the design of a task may address related quality issues,
e.g., starting from input fields left empty.

\item \emph{Control execution}: Finally, some actions can be enacted
during runtime, that is, during the execution of a task while workers are
already working and submitting outputs. For example, if it is evident at
some point in time that not all workers will produce outputs, it could be 
an idea to re-deploy some tasks. 

\end{itemize}

Each of these strategies can be implemented using different,
concrete \emph{actions}. We discuss these in the following and
summarize the discussion in Appendix \ref{app:references} tabularly.

\subsection{Improve Data Quality}

\subsubsection{Cleanse data}
The precondition of any process for good quality in output is good quality 
in input. \citeN{Khazankin2012} point out that a crowdsourcing
platform cannot be responsible for the output quality if the input data 
is inaccurate, and the requester is responsible for the accuracy of the inputs.
In fact, workers may be reluctant to work on a task 
if they perceive the quality of the input (e.g., a blurred image) may
impact their likelihood of getting the reward \cite{DBLP:conf/amcis/SchulzeNS13}.
To overcome input quality problems, \citeN{bozzon2012answering}, for instance, 
propose specific data pre-processing steps to assemble, re-shape or filter
input data in CrowdSearcher. \citeN{Bigham:2010} use computer vision
techniques to improve the quality of pictures taken by blind people. Of
course, data cleansing can be applied also to output data.

\subsubsection{Aggregate outputs} In the book ``The Wisdom of Crowds,''
\citeN{crowdwisdombook} has shown with his ox weight guessing experiment 
that averaging (aggregating) the guesses of multiple people can produce an 
accurate overall guess. That is, adding redundancy (multiple workers working on a 
same task) and aggregating outputs can increase the quality 
of outputs, of course, at the cost of  paying more workers. The responses of 
workers can also be weighted based on their reliability to increase the influence 
of responses given by trusted and skilled workers \cite{aydin2014crowdsourcing}.
\rev{Particular attention is payed by literature to the aggregation of outputs in classification tasks \cite{Ho2016,Gao2016,Ok2016,Chao2012}, including Boolean ones \cite{de2015reliable,ukkonen2015crowdsourced}}


\subsubsection{Filter outputs} The assessment methods discussed in the previous section aim to tell apart ``good'' from ``bad'' items (outputs, workers, requesters, etc.). 
The corresponding assurance action that aims to improve quality is filtering 
out the bad items, so as to keep only the good ones. Filtering is very prominent
in crowdsourcing. For instance, \cite{Dow2012} filter outputs based on self- and
expert reviews, \citeN{hansen2013quality} filter outputs based on output 
agreements (with/without arbitration) and peer review, \citeN{Rao2013}
use majority votes to filter outputs, others use ground truth comparisons 
\cite{marcus2012counting}, and so on. \citeN{Jung2012} filter workers based
on their past performance. \rev{As an extension of the pure filtering of outputs,
these may further be cleaned of possible biases by workers, e.g., through
active learning \cite{Zhuang2015,Wauthier2011}.}

\subsubsection{Iterative improvement} Instead of asking one worker to evaluate
the work of another worker (assessment), it is also possible to ask the former to
directly improve the work of the latter (assurance). Multiple iterations of improvement
are of course possible. \citeN{little2010turkit}, for example, let workers iterate
over writing tasks to incrementally improve a text written collaboratively by the crowd,
while in \cite{little2010exploring} they apply iterative improvement to decipher 
blurred text. In Turkomatic \cite{Kulkarni2012}, workers can iteratively
split tasks down till they are manageable, thereby reducing the complexity of the
overall task. 


\subsection{Select People}


\subsubsection{Filter workers}
Similar to outputs, a requester may also filter workers to make sure they have 
the necessary skills, preparation, attitude, or similar for a task. This filtering can, 
for instance, be based on the worker profiles and look at skills and expertise
\cite{Allahbakhsh2013,zhao2013transfer}, badges (CrowdFlower), or
demographics (as a proxy for skills) \cite{kazai2012face}. \citeN{kazai2012face}
filter workers also by personality, while \citeN{li2014wisdom} look at worker
reliability, and \citeN{trustcol12} look at the reputation of workers inside the 
crowdsourcing platform. \citeN{hara2013combining} use statistical filtering 
(outlier elimination) to discard low-quality workers. \rev{\citeN{Abraham2016}, too,
use reputation to filter workers, but also propose the use of adaptive stopping
rules to lower the cost of tasks.} \rev{\citeN{Nushi2015} specifically filter workers
to increase crowd diversity.}


\subsubsection{Reject workers}
Instead of selecting good workers, it is also possible to skim out (reject) non-fitting, 
fake, malicious or colluding workers, often also called attackers or cheaters.
Typical adversarial techniques are randomly posting answers, artificially generating
answers, or duplicating them from other sources \cite{difallah2012mechanical}
as well as collaborations among workers aimed at tricking the system (collusions).
Bots or software agents can be eliminated using CAPTCHAs \cite{lasecki2014information}.
\citeN{difallah2012mechanical} use start time and end time to predict if a given 
result comes from a cheater or not. It is important to properly tune the applied
thresholds to detect cheaters without affecting good workers \cite{Bozzon2013}. 
\citeN{marcus2012counting} compare ground truth data with task outputs to prevent
coordinated attacks from workers. \citeN{Allahbakhsh2014SOCA} identify collusions
from past collaborations of workers. 

\subsubsection{Assign workers}
Instead of waiting for tasks to be pulled by random workers, it may be more
effective to pro-actively push tasks to cherry-picked workers. If both the skills 
required by a task and the ones possessed by workers are defined, tasks can be
automatically assigned. For example, Mobileworks uses priority queues to assign
tasks \cite{kulkarni2012b}. \citeN{Allahbakhsh2013} assign stronger workers to 
harder tasks, \rev{while \citeN{roy2015task} specifically focus on knowledge-intensive
tasks and show the benefits of maintaining a pre-computed index for efficient
worker-to-task assignment.}  \citeN{difallah2013pick} show that
matches can also be derived from social network activity,
while \citeN{Kulkarni2014} identify experts in social networks and recruit 
them. If no skills regarding tasks and workers are given, a two-phase 
exploration-exploitation assignment algorithm can be applied \cite{ho2012online}. 
\rev{More dynamic approaches may even learn task assignment policies at 
runtime, e.g., to maximize the expected information gain \cite{Kobren2015}.}


\subsubsection{Recommend tasks}
Instead of assigning tasks to workers, identified task-worker matches can
also be used to provide workers with recommendations of tasks they very
likely are interested in. This still allows workers to decide to work on a
recommended task or not. A recommendation can be delivered, for example, as 
an email notification when a new task is posted on a platform
\cite{Bernstein201OptRealtime}, 
and subscriptions can be created using \url{www.turkalert.com}. 
Recommendations can be based on a worker's task browsing history
\cite{DBLP:journals/npl/YuenKL15}, or they may take into account implicit 
negative feedback \cite{lin2014signals}.

\subsubsection{Promote tasks}
With promoting a task we refer to all those actions that aim to enlarge the 
worker basis for tasks, even outside crowdsourcing platforms. More workers
means more diversity, tasks completed, speed, and similar. \citeN{hu2012deploying},
for instance, place widgets with integrated tasks on third-party services, such as
the International Children's Digital Library. \citeN{kulkarni2012b} ask workers 
to recruit others for a given task, others incentivize this kind of recruitment
\cite{nath2012threats}. Also posting task information on worker community 
sites, such as the ``HITsWorthTurkingFor'' reddit page, can increase the task's 
workforce. \rev{A more direct way of promoting tasks is to invite people, e.g.,
experts from online communities, to participate in tasks \cite{Oosterman2016}.}

\subsubsection{Situated crowdsourcing}
This means bringing tasks to the workers physically, where it
is more likely to encounter target workers, instead of waiting for 
them to join a crowdsourcing platform and to actively look for work. 
For instance, dedicated kiosks installed next to a library entrance 
\cite{hosio2014situated}, vending machines installed in front of the 
major lecture hall in a CS building \cite{Heimerl:2012:CEL:2207676.2208619},
\rev{and public displays \cite{Niforatos2016}} have been used for the 
crowdsourcing of small tasks. Similarly, \citeN{vaish2014twitch}
ask mobile phone users for micro-contributions each time they unlock their 
phone, exploiting the common habit of turning to the mobile phone in spare moments.

\subsubsection{Recruit teams}
Team-based recruitment approaches overcome the problem of recruiting 
enough workers by finding and recruiting teams of workers who have profiles 
matching the task requirements. Forming and recruiting teams is possible in 
smaller communities such as communities that are specialized in specific services like
IT technical or business services \cite{vukovic2010peoplecloud,schall2012expert}.
\citeN{retelny2014expert} identify experts in the crowd and organize them into
teams to solve complex tasks. \citeN{li2014wisdom} run trial tasks to discover 
groups of workers producing higher quality than average and target them for 
future work on the same task. \rev{\citeN{Rokicki2015} study different team 
competition designs to improve performance.}

\subsection{Improve Extrinsic Motivation}

\subsubsection{Tailor rewards}
One of the key properties of each task is the reward. Choosing the right form 
and/or amount of the reward is thus of utmost importance for the success of a
crowdsourced task. \citeN{faradani2011s} \rev{and \citeN{Ho2015}} show that it is 
important to tweak the amount of the reward properly, so as to obtain good results.
\rev{\citeN{radanovic2016learning} demonstrate the effectiveness 
of dynamically adjusting payments based on output quality.}
\citeN{DBLP:conf/hcomp/MaoKCHSLS13} study the effectiveness
of different rewarding schemas, such as volunteering, pay per time, pay per task, 
pay per each data unit in a task, and show that workers are sensitive to 
the schemas. 
\rev{Along the same line, \citeN{Ikeda2016} show that paying tasks in bulks may
increase the task completion rate, while coupons or material goods decrease 
participation.}
\citeN{scekic2013incentives} also discuss deferred compensation
and relative evaluation as well as team-based compensation schemas. 
\citeN{Sakurai2013} propose reward strategies based on worker performance.
\citeN{singer2013pricing} study how to maximize the amount of tasks completed 
with a given budget using different pricing strategies.
\citeN{rokicki2014competitive} study gambling-based rewarding strategies.
\rev{Of course, also non-monetary rewards, such as badges, can be are seen as
``virtual trophies'' \cite{scekic2013incentives} that motivate workers, yet 
\citeN{Kobren2015} also show that these kinds of objectives must be designed
carefully otherwise they may produce detrimental effects.}

\subsubsection{Pay bonus}
A bonus is an additional, discretionary reward for good performance added to 
the base reward of a task. Bonuses are typically granted for the achievement of 
predefined goals or for reaching the threshold of key performance indicators (KPIs)
\cite{scekic2013incentives}. \citeN{Difallah2014} grant bonuses in response to
workers meeting given milestones. \citeN{Yin2014} top-up the base reward for
tasks if workers provide correct answers and if they react in less than 1 second. 
\citeN{yu2014comparison} give credits as base reward, bonuses of 5 cents
for additional tasks performed and of 10 US dollars for earning most credits 
(assigned every other month).
\citeN{Faltings2014} show that game-theoretic bonus schemas can also help 
eliminate worker bias, next to incentivizing them to work better.


\subsubsection{Promote workers}
Promoting a worker means raising the worker to a higher position compared to
his/her current one or compared to others. Promotions are particularly 
suitable to those environments that are characterized by a long-lasting engagement 
of workers, e.g., crowdsourcing environment with deferred payment schemas.
It has been shown that the prospect of a promotion, e.g., to get higher rewards 
or to obtain access to new types of tasks, increases motivation, also over longer
periods of time \cite{scekic2013incentives}. For example, \citeN{Dow2012} promote
workers from content producers to feedback providers (assessors), while 
\citeN{scekic2013programming}, next to promotions, also introduce the idea of
punishments.

\subsection{Improve Intrinsic Motivation}

\subsubsection{Share purpose}
Tasks that have a purpose that goes beyond the individual micro-task, that workers
understand and can identify with can help attract crowds to perform tasks with 
higher motivation, also for free by volunteering \cite{dontcheva2014combining}.
Examples of crowdsourcing initiatives driven by this sense of purpose are Zoouniverse
(\url{www.zooniverse.org}) described in \cite{DBLP:conf/hcomp/MaoKCHSLS13} or Wikipedia. 
As workers contribute for the purpose, rather than for the monetary reward (if any), 
these tasks are typically less attractive to spammers or adversarial workers.
\rev{\citeN{Kobayashi2015} show that tasks with an explicit social purpose may
help attract especially senior citizens as workers.} \rev{\citeN{Kaufman2016} identified that people tend to contribute more to volunteering tasks with few other contributors involved.}

\subsubsection{Self-monitoring}
Enabling workers to compare their performance with that of other workers 
(self-assessment) can switch workers into a competition mode that pushes them 
to perform better \cite{ipeirotis2014quizz,scekic2013incentives}. Most crowdsourcing
platforms today already assign workers a reliability or performance rating visible 
in the platform. \citeN{ipeirotis2014quizz} study the benefits of displaying scores 
to workers individually, overall crowd performance, and leaderboards with complete
or partial rankings of workers. Leaderboards have been used extensively so 
far \cite{rokicki2014competitive,dontcheva2014combining,preist2014competing}.

\subsubsection{Social transparency}
Social transparency means sharing identity (e.g., real name, persistent identity in
applications), content (e.g., comments), actions (e.g., tasks performed, endorsements)
among workers \cite{huang2013don}. Through transparency, workers build trust 
and bond with their co-workers and the requester and define standards and quality
control mechanisms that eventually may improve performance and output quality
\cite{huang2013don,viegas2007hidden}. According to \citeN{yu2014comparison},
maintaining good relationships between workers helps also to attract more workers.
\rev{In collaborative crowdsourcing scenarios, social pressure among workers (e.g.,
asking to stay) can have positive impacts on performance \cite{Feyisetan2016}.}

\subsubsection{\rev{Gamify task}}
\rev{Luis von Ahn introduced Games With A Purpose, where participants perform 
tasks in games for joy and entertainment, rather than financial 
reward \cite{Ahn2006}. \citeN{KrauseK15} found that for complex tasks workers produce better results in a gamified than in a paid condition; for simple tasks there 
was no difference. Designing tasks that induce curiosity is a task-agnostic strategy to improve worker retention \cite{Law2016}. \citeN{Feyisetan2015} propose a model to predict optimal combinations of financial and gamified incentives.}

\subsection{Train People}

\subsubsection{Prime workers}
Priming uses implicit mechanisms to induce observable changes in behavior that
are usually outside of the conscious awareness of the primed person \cite{Morris2012}.
Images, text, music and videos can be used to induce positive emotions in 
workers that in turn result in a positive effect on task performance, e.g., an image 
of a laughing child may induce workers to perform better \cite{Morris2012}.
\citeN {alagarai2014cognitively} use priming to help workers remember information. 
\citeN{Faltings2014} study the effectiveness of priming on task performance.

\subsubsection{Teach workers}
Teaching means providing workers with suitable instructions in order to enable them
to perform tasks. \rev{\citeN{Doroudi2016} show that providing expert examples and 
letting workers validate others' contributions are effective ways of teaching}. 
\rev{\citeN{Singla2014} use an algorithm to select expert examples to show to users}. 
To many workers, gaining or improving skills is a motivation
per se. This motivation can be enforced by in-person training like in Samasource 
(\url{http://www.samasource.org/}) or Mobileworks (\url{https://www.mobileworks.com/})
or through interactive tutorials, such as the ones described by
 \citeN{dontcheva2014combining}. Also, designing tasks in a way that helps workers 
learn improves output quality \cite{yu2014comparison}.

\subsubsection{Provide feedback}
Workers getting feedback from requesters about their performance provide results of better quality \cite{Dow2012}. The process of reviewing others' work itself improves quality \cite{yu2014comparison} and helps workers gain new skills \cite{zhu2014reviewing}. \citeN{Kulkarni2012} show how requester feedback is important to handle complex work in their Turkomatic platform.

\subsubsection{Team work}
Team working means working together on a same task, where ``together'' means
by interacting with each other (we don't consider it a team work if several workers
jointly label a set of images without any inter-personal communication). 
\citeN{kittur2010crowdsourcing}, for instance, use team work
for the translation of a poem, which requires negotiation between workers.
\citeN{andre2014effects} provide workers with a shared, collaborative 
editing space for creative writing tasks. \citeN{Dorn2012} propose flexible 
workflows to organize teams of collaborating workers.

\subsection{Improve Task Design}

\subsubsection{Lower complexity} 
One of the challenges in crowdsourcing is identifying the right granularity,
that is, complexity, for tasks. As \citeN{rogstadius2011assessment} show, 
the accuracy of outputs typically decreases with increasing task complexity.
From a design perspective, it is further important to implement tasks in a way
that limits cognitive complexity; for instance, comparing two objects is easier 
than identifying features of individual objects \cite{anderton2013analysis}. \rev{While the simplification of more complex tasks introduces longer completion times, it leads to higher quality; simpler tasks suit better if workers perform with interruptions \cite{Cheng2015b}.}

\subsubsection{Decompose task}
Another way to lower the complexity of a task is to decompose it into sub-tasks.
\citeN{Kittur2011} propose a partition-map-reduce approach in CrowdForge, 
where work is split into a set of sub-tasks executed in parallel and merged 
back later. \citeN{Kulkarni2012} propose a price-divide-solve algorithm in
Turkomatic, where workers themselves can decide whether to work on a task
themselves or rather split it into sub-tasks and then merge the outputs produced
by others.

\subsubsection{Separate duties}
Separation of duties is a design pattern lent from the business practice that 
requires multiple people to be involved in a decision, so as to prevent fraud 
or errors. Organizing work in such a way that workers have only one single 
task to perform may further lead to higher quality of outputs. 
\citeN{bernstein2010soylent}, for instance, propose a find-fix-verify approach
for text proofreading, where some workers identify errors
in a text (find), some others correct identified errors (fix), and others again 
check if there are no further mistakes left (verify). In \cite{Kulkarni2012}, 
workers decide if they prefer to split a task or to perform it.

\subsubsection{Validate worker inputs}
Validating worker inputs means checking that the values provided by the workers
through the form fields in the task UI comply with formal requirements,
e.g., no empty fields or only correct email addresses. This is a common design 
guideline that is, however, often not followed by requesters, eventually leading 
to low output quality. In Deluge \cite{bragg2013crowdsourcing} tasks are designed in such a way that workers must select at least one item from a list of 
available options. On CrowdFlower and AskSheet \cite{quinn2014asksheet}, the
requester can select allowed formats (e.g., an email, US address, phone number 
or even a custom regular expression) for fields in the task form and identify 
mandatory and optional input fields.


\subsubsection{Improve usability}
Just like for any kind of software, it is important that task UIs are usable and
properly follow common usability guidelines
\cite{nielsen2002homepage,khanna2010evaluating}.
Well-designed tasks lead to higher quality of the outputs \cite{Kazai2011}. 
For instance, it is a good practice to provide workers with clear and meaningful 
examples of good work \cite{willett2012strategies}.  Highlighting input fields 
adequately and placing them closer to the relevant content reduces search time 
and working memory load \cite{alagarai2014cognitively}. Designing tasks in a 
way that it takes the same or less time to perform a task properly rather than 
to cheat also helps avoid spammers \cite{userstudies}.

\subsubsection{\rev{Prompt for rationale}}
\rev{Collecting rationales from workers for their own work is a good way to 
encourage workers to be more conscious and to collect verifiable results, 
especially for subjective tasks \cite{Mcdonnell16}. \citeN{drapeau2016microtalk} 
suggest that allowing workers to adjust their work based on rationales provided 
by other workers may improve quality further.}

\subsubsection{\rev{Introduce breaks}}
\rev{Performing long sequences of monotonous tasks can be downing for workers. 
\citeN{Dai2015} show that introducing occasional breaks, such as playing games 
or reading a comic, may help increase workers' retention.
}

\subsubsection{\rev{Embrace error}}
\rev{There are classes of tasks where fast completion has higher priority than the quality of each individual worker's judgment. A way to approach such tasks is to design them encouraging workers to perform very fast, accepting and even embracing possible errors, which can be later rectified through suitable post-processing \cite{Krishna2016}.
}

\subsection{Control Execution}

\subsubsection{Reserve workers}
Maintaining a pool of workers ready to work on a task is an effective approach
to minimize waiting time that is especially useful if results are to be collected fast,
e.g., in real-time. For instance, it is possible to pay workers for the time they spend 
waiting for tasks \cite{Bernstein201OptRealtime}. To make sure workers stay focused
during the waiting time they can be asked to play a game \cite{lasecki2013real}. 
In case of requesters launching tasks only occasionally, it can be financially efficient 
to maintain a single pool of workers for multiple requesters. \citeN{Bigham:2010} adopt
this approach and notify workers of the retainment pool when a new task is published.

\subsubsection{Flood task list}
\citeN{Chilton:2010:TSH:1837885.1837889} have shown that workers tend to give
more attention to newer tasks when looking for tasks to work for. If a task does not
appear on the top of the first page of the task list or even goes to the second page 
of the listing, the attention devoted to it drops dramatically. To keep the attention of
workers high, \citeN{Bernstein201OptRealtime} have shown that repeatedly posting
tasks (flooding) inside a crowdsourcing platform indeed increases the task's 
visibility and attractiveness.

\subsubsection{Dynamically instantiate tasks}
Monitoring the work of the crowd can allow the requester to identify quality
issues while a task is still in execution, e.g., workers not completing their work or doing so with too low level of quality. \citeN{KucherbaevCSCW2016} show
how dynamic re-launching of tasks, that are automatically identified as abandoned, helps to lower overall task execution times in exchange of a small cost overhead. \citeN{Bozzon2013} support the dynamic re-planning of task deployments. \rev{\citeN{yan2011active} study how to actively learn to choose workers to minimize speed and/or time. Many study how to maximize quality by dynamically instantiating and assigning work under budget constraints \cite{Li2016,Tran-ThanhHRRJ15,chen13a,Karger2011,Karger2014}. \citeN{Bansal2016} use content similarity to dynamically identify data items to label and propagate labels to similar items.}




\subsubsection{Control task order}
If a group of tasks presents interdependencies where the output of one task
affects the usefulness of another task, controlling the order of task deployment 
can help avoid useless extra costs. For instance, in a set of comparison tasks if 
$a = b$ and $b \neq c$, then there is no reason to compare also $a$ and $c$ 
\cite{vesdapunt2014crowdsourcing}. \citeN{marcus2012counting} show how to
use selectivity estimations used in traditional databases for query optimization 
to order tasks and reduce them in number. 
\rev{\citeN{Lasecki2015} and \citeN{Newell2016} suggest to group tasks with 
related content into batches, because tasks that have already been completed 
by a worker affect his/her focus in subsequent tasks, and the too high a diversity 
of task content inside a same batch of tasks leads to interruptions due to context 
switch. Yet, \citeN{Eickhoff2013} also show that large 
batches tend to attract more cheaters than small batches.
 \citeN{Difallah2016} address context switch and minimize latency using 
scheduling techniques guaranteeing that all tasks get equal attention.}

\subsubsection{Inter-task coordination}
More complex tasks, especially composite tasks that involve multiple different
sub-tasks, can be managed by automating the respective crowdsourcing workflow.
\citeN{Bozzon2013} propose an event-condition-action approach to organize
tasks. \citeN{KucherbaevIC2015} overview workflow automation instruments 
tailored to crowdsourcing, e.g. Turkit \cite{little2010turkit} supporting scripting and CrowdLang \cite{minder2012crowdlang} supporting visual modeling.

%% file: sections/8.soa.tex
\section{Analysis of State-Of-The-Art Crowdsourcing Platforms}
\label{sec:analysis}

In the following, we discuss and compare a selection of 
state-of-the-art crowdsourcing platforms
with the help of the taxonomy introduced in 
this article. We thus specifically look at the quality 
model/attributes, the assessment methods, and the 
assurance actions supported by the approaches.

The crowdsourcing platforms we analyze represent a 
selection of heterogeneous instruments drawn from both industrial systems and academic research
prototypes. The selection is by no means intended to be 
complete, nor does it represent a list of ``most popular''
instruments. It is rather the result of an internal discussion 
among the authors of this article of the platforms we found 
in our research, the platforms we looked at in the analysis,
and the platforms we personally worked with over the last
years. The goal of the selection was to provide a varied
picture of the platforms that characterize today's crowdsourcing
landscape.

The result of this discussion is the selection of the following
14 platforms:
\emph{Mechanical Turk} (\url{http://www.mturk.com}), 
one of the first crowdsourcing platforms for paid micro-tasks;
\emph{CrowdFlower} (\url{http://www.crowdflower.com}),
a meta-platform for micro-tasks that acts as proxy toward other 
platforms;
\emph{MobileWorks} (\url{http://www.mobileworks.com}),
a platform with an ethical mission that pays workers hourly
wages;
\emph{Crowdcrafting} (\url{http://crowdcrafting.org}), which 
targets scientists and non-paid volunteers;
\emph{Turkit} \cite{Little2010}, a JavaScript-based language 
for the programmatic coordination and deployment of tasks on 
Mechanical Turk;
\emph{Jabberwocky} \cite{Ahmad2011}, a MapReduce-based 
human computation framework with an own programming language;
\emph{CrowdWeaver} \cite{kittur2012crowdweaver}, a model-based 
tool with a proprietary notation and crowdsourcing-specific constructs; 
\emph{AskSheet} \cite{quinn2014asksheet}, a Google Spreadsheet 
extension with functions for the integration of crowdsourcing tasks; 
\emph{Turkomatic} \cite{Kulkarni2012}, a crowdsourcing tool that 
delegates not only work to the crowd but also task management 
operations (e.g., splitting tasks);
\emph{CrowdForge} \cite{Kittur2011}, a crowdsourcing framework 
similar to Turkomatic that follows the Partition-Map-Reduce approach;
\emph{Upwork} (\url{https://www.upwork.com/}), an auction-based 
platform for freelancers in different domains (e.g., software development
or writing);
\emph{99designs} (\url{http://99designs.it/}), a contest-based 
platform for graphical design freelancers;
\emph{Topcoder} (\url{https://www.topcoder.com/}), a contest-based 
platform for software developers and designers;
\emph{Innocentive} (\url{http://www.innocentive.com/}), a platform 
for the crowdsourcing of innovation challenges.

\begin{wrapfigure}[15]{r}{0.35\textwidth}
\begin{center}
\vspace{-8mm}
\includegraphics[width = 0.35\textwidth]{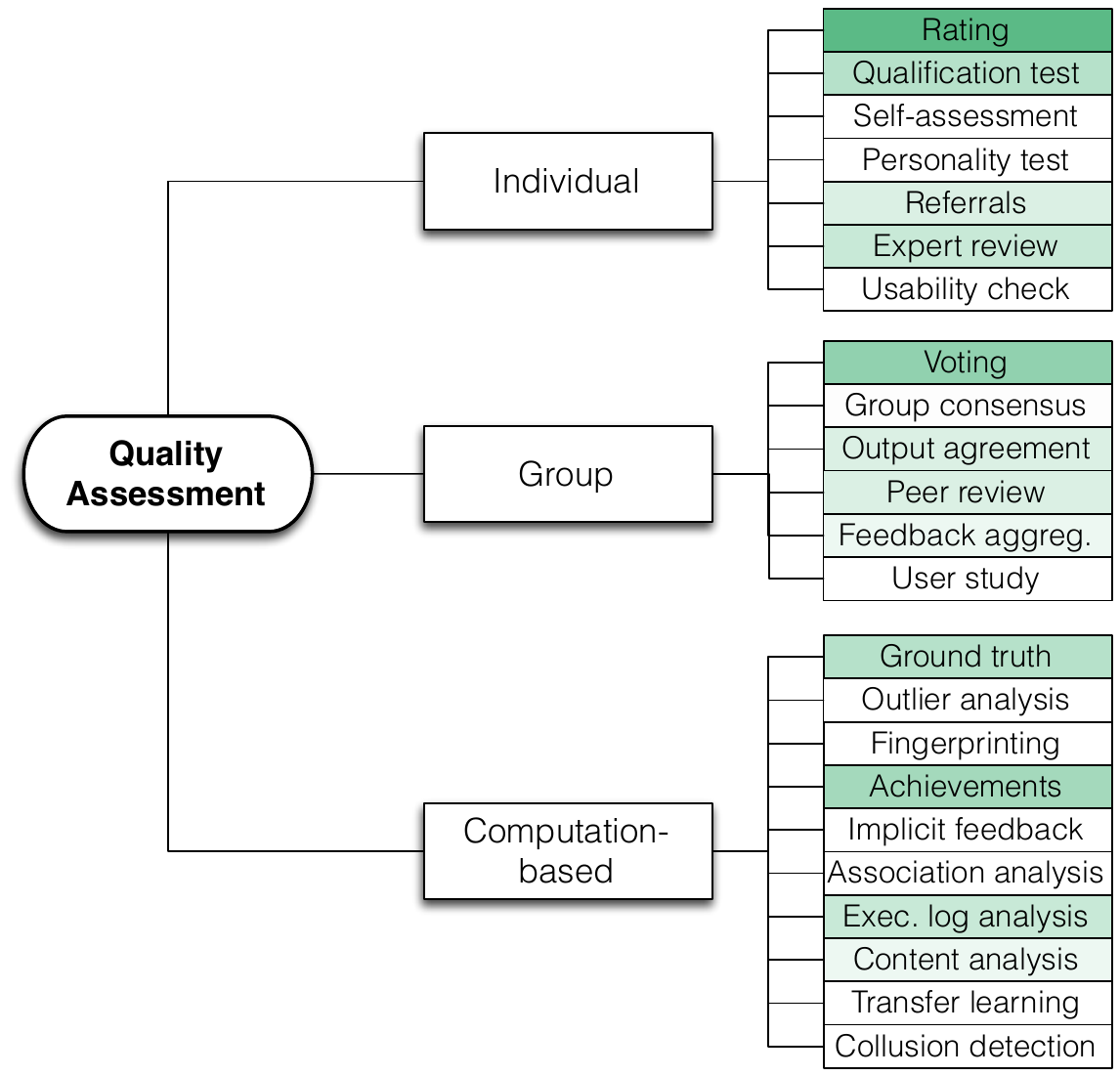}
\end{center}
\vspace{0mm}
\caption{\rev{Assessment heat map}}
\label{fig:heatassess}
\end{wrapfigure}

We use the taxonomy to classify each of these platforms or
tools individually in Table \ref{platforms-comparison} and to summarize
the supported quality attributes, assessment methods and 
assurance actions from a qualitative point of view. 
Figures \ref{fig:heatmodel}--\ref{fig:heatassure}
provide a quantitative summary of the table in the form of heat maps
that color each of the leaves (quality attributes, assessment methods
and assurance actions, respectively) of the Figures 
\ref{fig:qualitymodel}--\ref{fig:assurance} with a different intensity 
according to how many of the platforms/tools support the respective 
feature; possible values thus range from 0 (white) to 14 
(dark green). We discuss the findings in the following subsections.

{
\scriptsize
\begin{longtable}{|L{0.25cm}|L{4cm}|L{3cm}|L{5cm}|}
\caption{Crowdsourcing platforms comparison by supported quality model/attributes, assessment methods and assurance actions.} \label{platforms-comparison} \\
\hline
 & \textbf{Quality attributes} & \textbf{Assessment} & \textbf{Assurance} \endfirsthead
\hline
 & \textbf{Quality attributes} & \textbf{Assessment} & \textbf{Assurance} \endhead
\hline
\rot{\textbf{MTurk}} & data (accuracy), task incentives (extrinsic), task terms and conditions (privacy), task performance (cost efficiency, time efficiency), worker profile (location), worker credentials (skills), worker experience (badges, reliability) & rating (reliability), qualification tests (skills), ground truth questions (accuracy), rating-based achievements (badges) & filter workers by reliability (acceptance rate, tasks completed), by badges (domain specific master skill) or location; assign workers; reject workers; tailor rewards; pay bonus; provide feedback to workers; validate worker inputs; \rev{prompt for rationale}\\ 
\hline
\rot{\textbf{CrowdFlower}} & data (accuracy, timeliness), task description (clarity, complexity), task incentives (extrinsic), task terms and conditions (privacy, information security), task performance (cost efficiency, time efficiency), worker profile (location), worker credentials (skills), worker experience (badges, reliability) & rating (reliability), qualification tests (skills), output agreement (accuracy, reliability), feedback aggregation (task satisfaction survey on clarity, complexity, incentives), ground truth questions (accuracy), rating-based achievements (badges), execution log analysis (accuracy, timeliness) & aggregate outputs (consistency, accuracy), filter workers (by country, by distribution channel, by NDA), badges (obtainable level), skills (language), reject workers, tailor rewards; pay bonus, provide feedback, validate worker inputs, \rev{prompt for rationale}, control task order (select an order in which tasks are deployed), inter-task coordination (Crowdflower workflow plugin)\\ 
\hline
\rot{\textbf{MobileWorks}} & data (accuracy, timeliness), task incentives (extrinsic), task terms and conditions (privacy), worker profile (age, gender, location), worker credentials (skills), worker experiences (badges, reliability), group (availability) & rating (reliability), qualification tests (skills), expert review (accuracy and reliability of group members), peer review (accuracy), ground truth (accuracy) & filter workers, reject workers, assign workers, promote tasks (workers involve referrals), recruit teams (recruit local small teams, interview via Skype), tailor rewards, pay bonus, promote workers to team leaders, teach workers, provide feedback, teamwork, validate worker inputs, \rev{prompt for rationale}, control task order\\ 
\hline
\rot{\textbf{CrowdCrafting}} & intrinsic incentives (citizenscience), task terms and conditions (privacy), task performance (time efficiency), worker profile (location, personal details) & execution log analysis (time efficiency) & promote tasks (featuring on the platform), share purpose (tasks from high impact scientific fields), self-monitoring (contributions leaderboard), social transparency (optional public worker profiles), \rev{prompt for rationale}, control task order (tasks order priority, scheduling - depth first, breadth first, random)\\ 
\hline
\rot{\textbf{TurKit}} & data (accuracy, timeliness), extrinsic incentives (reward)  & voting (accuracy) & filter outputs, iterative improvement, tailor rewards, separate duties (explicit voting tasks), dyn. instantiate tasks (as a part of iterative improvement), control task order (via automatic workflow), inter-task coordination (programmatically using JavaScript-like scripts)\\ 
\hline
\rot{\textbf{Jabberwocky}} & data (accuracy), extrinsic incentives (reward), worker profile (age, gender, location, custom attributes), credentials (skills, certificates) & rating (accuracy), voting (accuracy) & aggregate outputs, filter outputs, iterative improvement, filter workers (rich profiles), assign workers, tailor rewards, inter-task coordination (Dog programs)\\ 
\hline
\rot{\textbf{CrowdWeaver}} & data (accuracy, timeliness), extrinsic incentives (reward), task performance (cost efficiency, time efficiency) & voting (accuracy), output agreement (accuracy), ground truth (accuracy) & cleanse data (divide, permute tasks), inter-task coordination (supports runtime workflow edits) \\
\hline
\rot{\textbf{AskSheet}} & data (accuracy), extrinsic incentives (reward), task performance (cost efficiency) & rating (accuracy), voting (accuracy) & cleanse data (facilitated through spreadsheet paradigm), aggregate outputs (spreadsheet formulas), filter outputs (spreadsheet formulas, worker vote), tailor rewards, validate worker inputs (enforce predefined bounds and types), dyn. instantiate tasks (launching extra instances until a certain threshold is reached), control task order (prioritization), inter-task coordination (using referral links in formulas)\\ 
\hline
\rot{\textbf{Turkomatic}} & data (accuracy), extrinsic incentives (reward), task description (complexity) & voting (accuracy) & aggregate outputs (via ``merging" tasks), filter outputs (via ``voting" tasks), tailor rewards, teamwork (via runtime workflow edits), decompose task (via ``subdivision" tasks), inter-task coordination (emerges at runtime following the price-divide-solve algorithm)\\ 
\hline
\rot{\textbf{CrowdForge}} & data (accuracy), extrinsic incentives (reward), task description (complexity) & voting (accuracy) & aggregate outputs (via ``reduce" step), filter outputs (via ``voting" tasks), decompose task (via dynamic partitioning), inter-task coordination (according to partition-map-reduce approach with possible nesting)\\ 
\hline
\rot{\textbf{Upwork}} & data (accuracy), extrinsic incentives (reward), task terms and conditions (IP), task performance (cost efficiency), requester (reputation), worker profile (location), worker credentials (skills, certificates, portfolio), experiences (badges, reliability) & rating (accuracy), qualification test (skills), referrals (reliability), expert review (accuracy), achievements (badges), execution log analysis (monitoring of worked hours) & filter workers (also via interviews), reject workers (disputes), assign workers (invite to work), recruit teams, tailor rewards (per hour vs. fixed price), pay bonus, social transparency, provide feedback, teamwork, \rev{prompt for rationale}\\ 
\hline
\rot{\textbf{99designs}} & data (accuracy, timeliness), extrinsic incentives (reward), task terms and conditions (IP), worker profile (location), worker credentials (skills), worker experiences (badges, reliability) & rating (accuracy, reliability), referrals (reliability), achievements (badges) & filter outputs (based on competition), filter workers (profile, experience), assign workers (invite to work), tailor rewards (predefined plans or direct negotiation), self-monitoring (submissions of others are optionally visible), social transparency (workers can use real identities and build professional profiles), provide feedback, teamwork, \rev{prompt for rationale}\\ 
\hline
\rot{\textbf{Topcoder}} & data (accuracy, timeliness), extrinsic incentives (reward), task terms and conditions (IP, information security), worker profile (location), worker credentials (skills), experiences (badges, reliability, custom performance metrics)  & rating (accuracy), peer review (called community review), achievements (badges), content analysis (unit tests on submitted code) & filter outputs (test-based or community review), tailor rewards, pay bonus, self-monitoring (leaderboard), social transparency (workers can link social network profiles with their identities), \rev{prompt for rationale}\\ 
\hline
\rot{\textbf{Innocentive}} & data (accuracy), extrinsic incentives (reward), task terms and conditions (IP, compliance), worker profile (location, rich profile), worker credentials (skills, certificates) & rating (accuracy), expert review (accuracy) & filter outputs (proposals filtered manually), recommend tasks (based on workers skills and interests), tailor rewards, \rev{prompt for rationale}\\ 
\hline
\end{longtable}
}


\subsection{Quality Model}

\begin{wrapfigure}[21]{r}{0.55\textwidth}
\begin{center}
\vspace{-8mm}
\includegraphics[width = 0.55\textwidth]{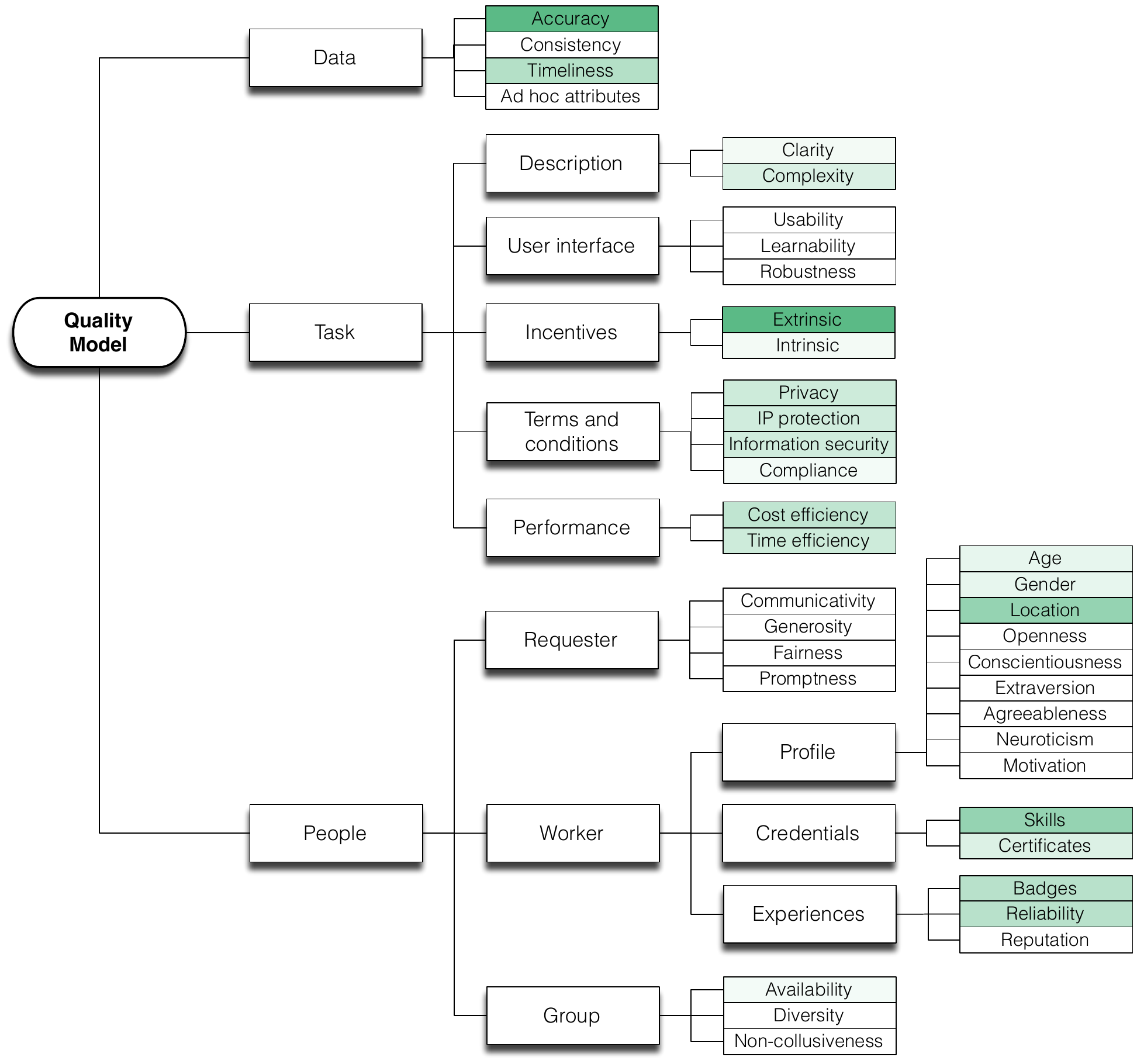}
\end{center}
\vspace{0mm}
\caption{\rev{Heat map of the quality model.}}
\label{fig:heatmodel}
\end{wrapfigure}

Figure \ref{fig:heatmodel} illustrates how many of the studied platforms
support each of the attributes of the quality model derived in this article.
Immediately, it is evident that the core concern almost all platforms (13
out of 14) have is the accuracy of the outputs produced by the
crowd. This is not surprising, as high quality outputs are one of the key
drivers of crowdsourcing in the first place (next to cost and time). In 
order to allow the requester to tweak quality, most platforms allow the 
requester to fine-tune the extrinsic incentives (rewards) given for tasks (13/14), to select
workers based on age (2/14), gender (2/14), location (9/14), skills (8/14).
In addition, approximately half of the platforms also implement proper
reliability tracking (6/14) or reputation management systems (6/14) for
worker selection.

In order to understand these numbers better, it is necessary to disaggregate 
them. For instance, it is important to note that worker profiles are typically
kept simple by the marketplace platforms (e.g., Mechanical Turk or 
CrowdFlower), while they are more sophisticated for auction-/contest-based
platforms that target skilled freelances (e.g., Topcoder or 99designs). In 
fact, the former (but also the research prototypes that target the 
coordination of multiple tasks on top of existing marketplace platforms) 
mostly focus on simple microtasks for which the problem
usually is finding workers at all, not finding the best ones possible; in the 
latter platforms, instead, the profile has an advertising purpose as well 
and therefore plays a more important role. All platforms generally protect
the privacy of their users; the research prototypes adopt the policies of the
underlying platform, while the platforms for freelancers may disclose personal
information to enable transactions among collaborating actors. Support for IP
protection comes in the form of NDAs or transfer agreements for freelancers.
Worker assessment seems mostly based on skills, reliability, badges and/or 
reputation. That is, crowdsourcing tends to be meritocratic. 

However, overall there is only little support for 
the different quality attributes identified in these articles. Mostly, this is
explained by the different focus of research (wide spectrum of attributes) 
and commercial practice (narrow focus). For both areas, we identify the
following points as possible future research directions:

\begin{itemize}

\item \emph{Personality.}
The character and behavior of workers and 
requesters, acknowledged by research as directly impacting the 
quality of outputs and the satisfaction of both, is commonly 
neglected by state-of-the-art instruments. Yet, there seems to be 
an opportunity for platforms that also aim to improve the attitude 
of people, e.g., by facilitating the creation of shared values, social
contacts, or social norms. People that feel well in their workplace
perform better and are more engaged in their work.
\item \emph{Transparency.}
In general, while workers on many platforms are anonymous to requesters, 
it is important to note that requesters are even more anonymous to workers.
In order to increase mutual trust, it is advisable that also requesters
be assessed properly and participate more actively in the actual work.
How this assessment and/or collaboration could happen is not straightforward
in a domain where each new task may put into communication 
completely new and unknown actors. 

\item \emph{Group work.}
While there are first attempts of organizing workers into groups and to
facilitate the collaboration between workers and requesters, the quality 
and benefit of group work is still not fully studied and understood. In this 
respect, we believe the attributes considered so far are not sufficient to
help characterize the quality of and leverage on the full power of the crowd.
But first of all, more support for group work and collaboration by the platforms 
themselves is needed.

\item \emph{User interface quality.}
Surprisingly, only very little attention is paid by the studied platforms to 
the quality of the user interface of tasks deployed by requesters. Some
platforms (e.g., CrowdFlower) provide crowdsourcing as a service to
their customers, whereby they also design and control the quality of the 
respective UIs. Yet, on the one hand, usability and learnability are still
concepts that have not percolated into concrete tools and, on the other 
hand, they are still too generic as attributes to really help requesters to
develop better interfaces.

\end{itemize}


\subsection{Quality Assessment}

Figure \ref{fig:heatassess} illustrates the support of the discussed
assessment methods. As in the case of the quality model, also here
we see that only about half of the methods identified previously are
also implemented by current crowdsourcing platforms. Rating (9 out 
of 14) and voting (6/14) are the most prominent assessment methods,
the former mostly used by requesters to assess work/workers, the 
latter mostly used by workers for the peer assessment of work/workers.
It is further evident that the current support of assessment methods
is mostly limited to technically simple methods, while more complex
capabilities are crowdsourced themselves. For instance, the group
assessment methods are well developed overall, while the 
computation-based methods are still limited.

Again, it is good to disaggregate the numbers. While rating is almost
the only feedback mechanism in the marketplace platforms, and it is
essential for the proper functioning of the platforms based on contests,
it is interesting to note that almost none of the research prototypes
makes use of rating. Instead, given their focus on the coordination of
tasks these platforms heavily leverage on voting for quality assessment,
an activity that naturally involves multiple workers and possibly the 
requester and external experts. In line with the limited user profiles
featured by typical marketplace platforms, they instead prominently propose
the use of qualification tests to select workers. On the contrary, the 
auction- and contest-based platforms bet on achievements as
an automated technique to assess performance (e.g., badges).

For the future, we identify the following challenges in terms of quality
assessment:

\begin{itemize}

\item \emph{Self-assessment.}
This assessment method is underestimated by current practice,
despite its proven benefit to learning and to the quality of produced
outputs. How to make self-assessment an integral and integrated 
part of crowdsourcing in general is however non trivial and still open.

\item \emph{User interface assessment.}
In line with our comment on the quality attributes regarding UI quality, 
also in terms of assessment methods there is huge room for improvement.
Significant effort still needs to be invested into the development of proper 
guidelines for the design of intuitive and robust task UIs, as well as into
automated methods of their assessment, e.g., as attempted by 
\citeN{Miniukovich:2015}.

\item \emph{Runtime analytics.}
Assessment methods are mostly applied after task execution, while they 
could easily be applied also during task execution to enable preemptive 
interventions aimed at increasing quality while a task is being worked on.
Suitable analytics features and interventions could not only improve the 
accuracy of outputs, but also their timeliness and cost efficiency.

\end{itemize}

\begin{wrapfigure}[20]{r}{0.44\textwidth}
\begin{center}
\vspace{-8mm}
\includegraphics[width = 0.44\textwidth]{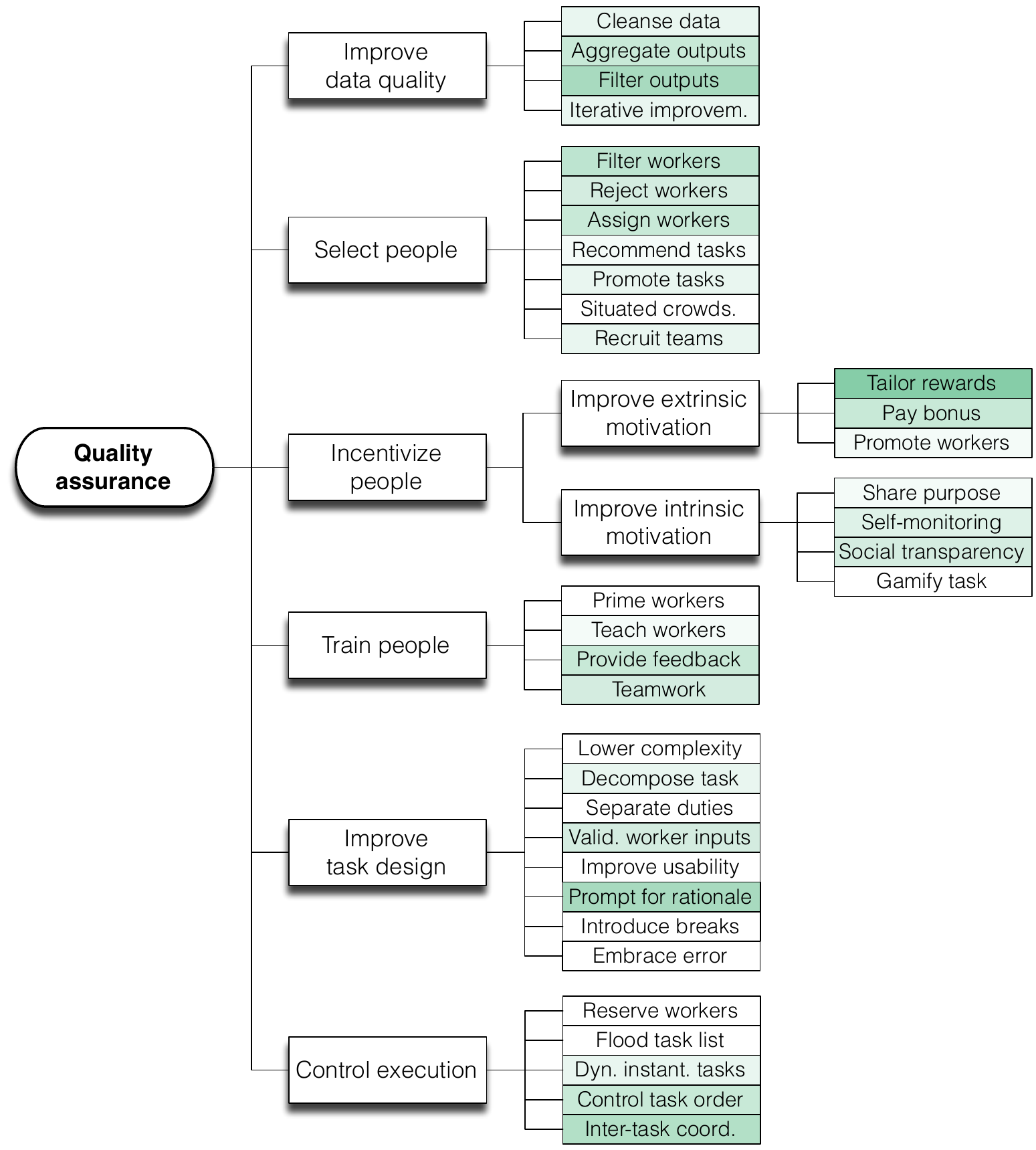}
\end{center}
\vspace{0mm}
\caption{\rev{Heat map of assurance model}}
\label{fig:heatassure}
\end{wrapfigure}

\subsection{Quality Assurance}

Finally, Figure \ref{fig:heatassure} illustrates the state of the art
in quality assurance. It is surprising to see that the spectrum of
assurance actions introduced earlier in this articles is explored
almost completely by current crowdsourcing platforms. Of course,
the tailoring of the reward (11 out of 14) dominates, but also filter
outputs (8/14) and filter workers (6/14) are adopted relatively widely.
\rev{Prompting for rationals is supported where requesters can
design task input forms or interact with workers (8/14).} 
Only situated crowdsourcing, prime 
workers, improve usability (out of the control of platforms), reserve workers,
and flood task list are not supported by any of the platforms.

The disaggregation of the data reveals that the strong support of
inter-task coordination (7/14) and task order control (5/14) mostly stems 
from the research prototypes that specifically focus on this aspect. 
To the best of 
our knowledge, only CrowdFlower internally uses a
workflow engine for task automation, and only Innocentive
seems to make use of task recommendations. Interestingly, it
appears that the research community with its prototypes for task
automation mostly concentrates on assuring quality by promoting
actions with effects that are limited to a given context only, e.g.,
controlling the order of tasks or splitting tasks into smaller chunks.
Commercial platforms rather look at more comprehensive actions,
such as intrinsic motivators (e.g, social transparency) or team building,
which have effects that cross tasks and are longer lasting in nature.

We identify the following points as crucial aspects to approach next:

\begin{itemize}

\item \emph{Task recommendation.}
With the increase of the popularity of crowdsourcing, both the 
number of workers and that of tasks published on crowdsourcing
platforms is destined to grow. It is thus increasingly hard for workers
to find tasks they are interested in, capable of, and good at. This 
asks crowdsourcing platforms to help match workers and work, 
which means recommending tasks to workers with the right skills.
Research is very active in this area, but support for custom worker 
profiles and recommendation algorithms is still missing.

\item \emph{Long-term relationships.}
To make crowdsourcing sustainable in the longer term, it
may be necessary that requesters and workers establish more
tight relationships, e.g., to train workers with skills that
will serve them in the future and assure them predefined
levels of income. Continuous and task-specific training must 
turn into common practice and be seen as an investment by both
workers and requesters.

\item \emph{Workflow integration.}
Finally, with the increasing complexity of work that is crowdsourced,
the need for coordination and automation will increase too. So far,
the problem has been studied only in the form of isolated research 
prototypes. The challenge now is conceiving principles, techniques
and tools that enable the seamless integration of crowd workflows 
into existing IT and business practices.
\end{itemize}

%% file: sections/10.discussion.tex

\section{Conclusion and Outlook}
\label{sec:discussion}
By now, crowdsourcing is a well established practice and a concrete option
to solve problems that neither individuals nor computers may be able to solve 
on their own (nor together), while they can be solved by asking help to 
contributors that are not known but that can bring in their human intelligence.
With this survey, we comprehensively studied one of the key challenges of
crowdsourcing, i.e., quality control. We analyzed literature on crowdsourcing
published in major journals and conferences from 2009 onwards and synthesized
a quality model that represents all the attributes that have been studied so far
to understand quality in crowdsourcing. We accompanied the model with a 
comprehensive discussion of the methods and actions that have been used
to assess quality and to enforce quality, respectively. From the survey 
it is evident that, although quality control is perceived as crucial by all actors 
involved in crowdsourcing and significant effort has already been invested into
it, we are still far from a practice without quality issues that effectively delivers
human intelligence to its customers. 

We consider the following two areas for future work as particularly critical to 
guarantee quality and sustainability in the longer term: 

\begin{itemize}

\item \emph{Domain-specific services}. 
Most crowdsourcing platforms, especially micro-task platforms and research 
prototypes, still position themselves as technology providers managing
the crowd and tasks from an abstract point of view. Crowdsourcing a piece
of work thus requires requesters to possess intimate crowdsourcing expertise
(e.g., to control quality) and to ``program'' the crowd on their own. To make crowdsourcing
more accessible, domain-specific service providers
that know the domain requirements, tasks and concerns and that can 
effectively assist also less skilled requesters in all phases of the crowdsourcing 
process are needed. CrowdFlower, for example,
already positions itself as a Data Science platform; 
platforms for creative tasks like 99designs or Topcoder are domain-specific by design.
Research on quality control has produced significant contributions so far, but it too 
needs to focus on domain-specific aspects if it wants to excel. Guaranteeing the 
quality of labels for images is just so different from doing so, e.g., for text translations.


\item \emph{Crowd work regulation and ethics}. Crowdsourcing is a worldwide 
phenomenon and practice that allows workers and requesters from all over the 
world to establish business relationships. Yet, common rules and regulations for 
crowd work, e.g., regarding taxation, pension, superannuation, resolution of 
disputes and similar, are still missing. Each crowdsourcing provider provides its
own rules and legal protections so far, if at all. Similarly, the ecosystem as a whole
(including workers, requesters and platform providers alike) needs to grow shared 
work ethics and standards. MobileWorks, for instance, guarantees its workers hourly wages, but this can only represent a starting point for sustainable
crowd work. On the IT research side, controlling the compliance of rules, regulations
and ethical aspects may ask for novel monitoring and assessment techniques, e.g.,
in the context of collusion detection.

\end{itemize}

%

%

%% file: sections/A.references.tex

\section{References Maps}
\label{app:references}


\begin{table}[h]
\scriptsize
\setlength{\tabcolsep}{0.4em} 
\renewcommand{\arraystretch}{1.1}
\centering
\caption{Representative literature references for the identified quality attributes.}
\label{tab:modelliterature}
\begin{tabular}{lllll|p{9cm}|}
\cline{6-6}
 &  &  &  &  & \textbf{References} \\ \hline
\multicolumn{1}{|l|}{\multirow{43}{*}{\bf \rot{Quality Model}}} & \multicolumn{3}{c|}{\multirow{5}{*}{\bf \rot{Data}}} & Accuracy & \citeN{hansen2013quality}, \citeN{Kazai2011}, \citeN{Yin2014}, \citeN{Cao2014}, \citeN{Eickhoff2012}, \citeN{Kulkarni2012}, \citeN{Kern2010}   \\ \cline{5-6} 
\multicolumn{1}{|l|}{} & \multicolumn{3}{l|}{} & Consistency & \citeN{huang2013enhancing}, \citeN{Eickhoff2012} \\ \cline{5-6} 
\multicolumn{1}{|l|}{} & \multicolumn{3}{l|}{} & Timeliness & \citeN{kittur2013future}, \citeN{lasecki2013real}, \citeN{lasecki2013warping},  \citeN{Yin2014} \\ \cline{5-6} 
\multicolumn{1}{|l|}{} & \multicolumn{3}{l|}{} & Ad hoc attributes & \citeN{yu2014comparison}, \citeN{nguyen2014using} \\ \cline{2-6} 
\multicolumn{1}{|l|}{} & \multicolumn{1}{l|}{\multirow{13}{*}{\bf\rot{Task}}} & \multicolumn{2}{l|}{\multirow{3}{*}{\bf\rot{\begin{tabular}[c]{@{}l@{}}Task\\ desc.\end{tabular}}}} & Clarity & \citeN{Hossfeld2014}, \citeN{Tokarchuk2012}, \citeN{georgescu2012map}, \citeN{kulkarni2012b}  \\ \cline{5-6} 
\multicolumn{1}{|l|}{} & \multicolumn{1}{l|}{} & \multicolumn{2}{l|}{} & Complexity & \citeN{hu2012deploying}, \citeN{rogstadius2011assessment}, \citeN{difallah2013pick}  \\ \cline{3-6} 
\multicolumn{1}{|l|}{} & \multicolumn{1}{l|}{} & \multicolumn{2}{l|}{\multirow{4}{*}{\bf\rot{\begin{tabular}[c]{@{}l@{}}User\\ interf.\end{tabular}}}} & Usability & \citeN{khanna2010evaluating}, \citeN{noronha2011platemate}, \citeN{retelny2014expert}, \citeN{alagarai2014cognitively} \\ \cline{5-6} 
\multicolumn{1}{|l|}{} & \multicolumn{1}{l|}{} & \multicolumn{2}{l|}{} & Learnability & \citeN{willett2012strategies}, \citeN{heer2010crowdsourcing} \\ \cline{5-6} 
\multicolumn{1}{|l|}{} & \multicolumn{1}{l|}{} & \multicolumn{2}{l|}{} & Robustness & \citeN{Eickhoff2012}, \citeN{Hung2013} \\ \cline{3-6} 
\multicolumn{1}{|l|}{} & \multicolumn{1}{l|}{} & \multicolumn{2}{l|}{\multirow{2}{*}{\bf\rot{\begin{tabular}[c]{@{}l@{}}In-\\cent.\end{tabular}}}} & Extrinsic & \citeN{Hossfeld2014}, \citeN{singer2013pricing}, \citeN{Eickhoff2012} \\ \cline{5-6} 
\multicolumn{1}{|l|}{} & \multicolumn{1}{l|}{} & \multicolumn{2}{l|}{} & Intrinsic & \citeN{Hossfeld2014}, \citeN{Eickhoff2012} \\ \cline{3-6} 
\multicolumn{1}{|l|}{} & \multicolumn{1}{l|}{} & \multicolumn{2}{l|}{\multirow{4}{*}{\bf \rot{\begin{tabular}[c]{@{}l@{}}Terms \&\\ cond.\end{tabular}}}} & Privacy &  \citeN{lasecki2013real}, \rev{\citeN{Boutsis2016}} \\ \cline{5-6} 
\multicolumn{1}{|l|}{} & \multicolumn{1}{l|}{} & \multicolumn{2}{l|}{} & IP protection & \citeN{towardCS} \\ \cline{5-6} 
\multicolumn{1}{|l|}{} & \multicolumn{1}{l|}{} & \multicolumn{2}{l|}{} & Information security & \citeN{towardCS}, \rev{\citeN{Amor2016}} \\ \cline{5-6} 
\multicolumn{1}{|l|}{} & \multicolumn{1}{l|}{} & \multicolumn{2}{l|}{} & Compliance & \citeN{wolfson2011look}, \citeN{wang2012serf}, \citeN{irani2013turkopticon} \\ \cline{3-6} 
\multicolumn{1}{|l|}{} & \multicolumn{1}{l|}{} & \multicolumn{2}{l|}{\multirow{3}{*}{\bf \rot{Perf.}}} & Cost efficiency & \citeN{ipeirotis2014quizz}, \citeN{rokicki2014competitive},  \\ \cline{5-6} 
\multicolumn{1}{|l|}{} & \multicolumn{1}{l|}{} & \multicolumn{2}{l|}{} & Time efficiency & \citeN{Eickhoff2012}, \citeN{lin2014signals}, \citeN{hung2013evaluation}, \citeN{KucherbaevCSCW2016}, \rev{\citeN{Cheng2015a}}  \\ \cline{2-6} 
\multicolumn{1}{|l|}{} & \multicolumn{1}{l|}{\multirow{21}{*}{\bf \rot{People}}} & \multicolumn{2}{l|}{\multirow{4}{*}{\bf \rot{\begin{tabular}[c]{@{}l@{}}Reques-\\ ter.\end{tabular}}}} & Communicativity & \citeN{irani2013turkopticon} \\ \cline{5-6} 
\multicolumn{1}{|l|}{} & \multicolumn{1}{l|}{} & \multicolumn{2}{l|}{} & Generosity & \citeN{irani2013turkopticon} \\ \cline{5-6} 
\multicolumn{1}{|l|}{} & \multicolumn{1}{l|}{} & \multicolumn{2}{l|}{} & Fairness & \citeN{irani2013turkopticon}, \citeN{trustcol12} \\ \cline{5-6} 
\multicolumn{1}{|l|}{} & \multicolumn{1}{l|}{} & \multicolumn{2}{l|}{} & Promptness & \citeN{irani2013turkopticon} \\ \cline{3-6} 
\multicolumn{1}{|l|}{} & \multicolumn{1}{l|}{} & \multicolumn{1}{l|}{\multirow{14}{*}{\bf \rot{Worker}}} & \multicolumn{1}{l|}{\multirow{9}{*}{\bf \rot{Profile}}} & Age & \citeN{kazai2011worker} \\ \cline{5-6} 
\multicolumn{1}{|l|}{} & \multicolumn{1}{l|}{} & \multicolumn{1}{l|}{} & \multicolumn{1}{l|}{} & Gender & \citeN{kazai2011worker} \\ \cline{5-6} 
\multicolumn{1}{|l|}{} & \multicolumn{1}{l|}{} & \multicolumn{1}{l|}{} & \multicolumn{1}{l|}{} & Location & \citeN{kazai2011worker}, \citeN{Eickhoff2012}, \citeN{kazai2012face} \\ \cline{5-6} 
\multicolumn{1}{|l|}{} & \multicolumn{1}{l|}{} & \multicolumn{1}{l|}{} & \multicolumn{1}{l|}{} & Openness & \citeN{kazai2011worker}, \citeN{kazai2012face} \\ \cline{5-6} 
\multicolumn{1}{|l|}{} & \multicolumn{1}{l|}{} & \multicolumn{1}{l|}{} & \multicolumn{1}{l|}{} & Conscientiousness & \citeN{kazai2011worker}, \citeN{kazai2012face} \\ \cline{5-6} 
\multicolumn{1}{|l|}{} & \multicolumn{1}{l|}{} & \multicolumn{1}{l|}{} & \multicolumn{1}{l|}{} & Extraversion & \citeN{kazai2011worker}, \citeN{kazai2012face} \\ \cline{5-6} 
\multicolumn{1}{|l|}{} & \multicolumn{1}{l|}{} & \multicolumn{1}{l|}{} & \multicolumn{1}{l|}{} & Agreeableness & \citeN{kazai2011worker}, \citeN{kazai2012face} \\ \cline{5-6} 
\multicolumn{1}{|l|}{} & \multicolumn{1}{l|}{} & \multicolumn{1}{l|}{} & \multicolumn{1}{l|}{} & Neuroticism & \citeN{kazai2011worker}, \citeN{kazai2012face} \\ \cline{5-6} 
\multicolumn{1}{|l|}{} & \multicolumn{1}{l|}{} & \multicolumn{1}{l|}{} & \multicolumn{1}{l|}{} &\rev{ Motivation} & \rev{\citeN{Kobayashi2015}} \\ \cline{4-6} 
\multicolumn{1}{|l|}{} & \multicolumn{1}{l|}{} & \multicolumn{1}{l|}{} & \multicolumn{1}{l|}{\multirow{2}{*}{\bf \rot{Cr.}}} & Skills & \citeN{difallah2013pick}, \citeN{schall2014crowdsourcing}, \rev{\citeN{Mavridis2016}}  \\ \cline{5-6} 
\multicolumn{1}{|l|}{} & \multicolumn{1}{l|}{} & \multicolumn{1}{l|}{} & \multicolumn{1}{l|}{} & Certificates & \citeN{Allahbakhsh2013} \\ \cline{4-6} 
\multicolumn{1}{|l|}{} & \multicolumn{1}{l|}{} & \multicolumn{1}{l|}{} & \multicolumn{1}{l|}{\multirow{4}{*}{\bf \rot{Exp.}}} & Badges & \citeN{anderson2013steering}, \citeN{scekic2013incentives},  \\ \cline{5-6} 
\multicolumn{1}{|l|}{} & \multicolumn{1}{l|}{} & \multicolumn{1}{l|}{} & \multicolumn{1}{l|}{} & Reliability & \citeN{Kazai2011}, \citeN{dalvi2013aggregating}, \citeN{demartini2013large}, \citeN{Sakurai2013}, \rev{\citeN{Raykar2011}} \\ \cline{5-6} 
\multicolumn{1}{|l|}{} & \multicolumn{1}{l|}{} & \multicolumn{1}{l|}{} & \multicolumn{1}{l|}{} & Reputation & \citeN{Allahbakhsh2013}, \citeN{de2011reputation} \\ \cline{3-6} 
\multicolumn{1}{|l|}{} & \multicolumn{1}{l|}{} & \multicolumn{2}{c|}{\multirow{3}{*}{\bf \rot{Group}}} & Availability & \citeN{li2014wisdom}, \citeN{ambati2012}   \\ \cline{5-6} 
\multicolumn{1}{|l|}{} & \multicolumn{1}{l|}{} & \multicolumn{2}{l|}{} & Diversity & \citeN{crowdwisdombook}, \citeN{Livshits2014}, \citeN{willett2012strategies} \\ \cline{5-6} 
\multicolumn{1}{|l|}{} & \multicolumn{1}{l|}{} & \multicolumn{2}{l|}{} & Non-collusiveness & \citeN{KhudaBukhsh2014} \\ \hline
\end{tabular}
Perf. = Performance --- Cr. = Credentials --- Exp. = Experience --- Incent. = Incentives
\end{table}

%
%

\begin{table}[t]
\scriptsize
\setlength{\tabcolsep}{0.22em} 
\renewcommand{\arraystretch}{1.1}
\centering
\caption{Representative literature references for the identified assessment methods.}
\label{tab:assessmentliterature}
\begin{tabular}{lll|p{10.5cm}|}
\cline{4-4}
 &  &  &   {\bf References} \\ \hline 
\multicolumn{1}{|l|}{\multirow{23}{*}{{\bf \rot{Quality Assessment}}}} & 
\multicolumn{1}{l|}{\multirow{8}{*}{{\bf \rot{Individual}}}} & Rating &  \citeN{dalvi2013aggregating}, \citeN{yu2014comparison}, \citeN{nguyen2014using}, \citeN{Sakurai2013}, \citeN{irani2013turkopticon}, \rev{\citeN{Hata2017}, \citeN{Gaikwad2016}}   \\ \cline{3-4}
\multicolumn{1}{|l|}{} & \multicolumn{1}{l|}{} & Qualification test & \citeN{heer2010crowdsourcing}    \\ \cline{3-4} 
\multicolumn{1}{|l|}{} & \multicolumn{1}{l|}{} & Self-assessment &   \citeN{Dow2012}, \citeN{Sakurai2013}, \rev{\citeN{Shah2015}, \citeN{ShahZ16}}  \\ \cline{3-4}
\multicolumn{1}{|l|}{} & \multicolumn{1}{l|}{} & Personality test &  \citeN{kazai2011worker}, \citeN{kazai2012face}, \citeN{irani2013turkopticon} \\ \cline{3-4} 
\multicolumn{1}{|l|}{} & \multicolumn{1}{l|}{} & Referrals &  \citeN{bozzon2012answering} \\ \cline{3-4} 
\multicolumn{1}{|l|}{} & \multicolumn{1}{l|}{} & Expert review & \citeN{Dow2012}, \rev{\citeN{Hung2015}}   \\ \cline{3-4} 
\multicolumn{1}{|l|}{} & \multicolumn{1}{l|}{} & Usability check & \citeN{nielsen2002homepage}, \citeN{willett2012strategies}  \\ \cline{2-4} 
\multicolumn{1}{|l|}{} & \multicolumn{1}{l|}{\multirow{8}{*}{{\bf \rot{Group}}}} & Voting &  \citeN{Kulkarni2012}, \citeN{little2010turkit}, \citeN{caragiannis2014modal} ,\citeN{sun2012majority} \\ \cline{3-4}
\multicolumn{1}{|l|}{} & \multicolumn{1}{l|}{} & Group consensus & \citeN{Sheshadri2013}, \citeN{Eickhoff2012}, \rev{\citeN{ZhangTKDE2015}, \citeN{Kairam2016}}  \\ \cline{3-4} 
\multicolumn{1}{|l|}{} & \multicolumn{1}{l|}{} & Output agreement & \citeN{Waggoner2014}, \citeN{huang2013enhancing}, \rev{\citeN{Jagabathula2014}}  \\ \cline{3-4} 
\multicolumn{1}{|l|}{} & \multicolumn{1}{l|}{} & Peer review & \citeN{hansen2013quality}, \citeN{zhu2014reviewing}, \rev{\citeN{Whiting2017}} \\ \cline{3-4} 
\multicolumn{1}{|l|}{} & \multicolumn{1}{l|}{} & Feedback aggreg. &   \citeN{dalvi2013aggregating},  \citeN{trustcol12}, \citeN{iterativefilter1}, \citeN{iterativefilter2}, \citeN{AleksRep}, \citeN{hung2013evaluation}, \citeN{joglekar2013evaluating}, \rev{\citeN{Davtyan2015}} \\ \cline{3-4} 
\multicolumn{1}{|l|}{} & \multicolumn{1}{l|}{} & User study & \citeN{willett2012strategies}, \citeN{khanna2010evaluating}, \citeN{alagarai2014cognitively}  \\ \cline{2-4} 
\multicolumn{1}{|l|}{} & \multicolumn{1}{l|}{\multirow{12}{*}{{\bf \rot{Computation-based}}}} & Ground truth &  \citeN{huang2013enhancing}, \citeN{Eickhoff2012}, \citeN{hara2013combining}, \citeN{oleson2011programmatic}, \citeN{le2010ensuring}, \citeN{lasecki2014information}, \citeN{von2008recaptcha}, \rev{\citeN{Liu2013}, \citeN{ElMaarry2015}}  \\ \cline{3-4}
\multicolumn{1}{|l|}{} & \multicolumn{1}{l|}{} & Outlier analysis &  \citeN{rzeszotarski2012crowdscape}, \citeN{jung2011improving}, \citeN{marcus2012counting}  \\ \cline{3-4} 
\multicolumn{1}{|l|}{} & \multicolumn{1}{l|}{} & Fingerprinting & \citeN{rzeszotarski2011instrumenting}, \citeN{rzeszotarski2012crowdscape}, \rev{\citeN{Kazai2016}}  \\ \cline{3-4} 
\multicolumn{1}{|l|}{} & \multicolumn{1}{l|}{} & Achievements & \citeN{Massung2013}  \\ \cline{3-4} 
\multicolumn{1}{|l|}{} & \multicolumn{1}{l|}{} & Implicit feedback & \citeN{adler2007content}, \citeN{adler2011wikipedia}, \citeN{lin2014signals}, \citeN{difallah2013pick}  \\ \cline{3-4} 
\multicolumn{1}{|l|}{} & \multicolumn{1}{l|}{} & Association analysis & \citeN{RajasekharanMN13},   \\ \cline{3-4} 
\multicolumn{1}{|l|}{} & \multicolumn{1}{l|}{} & Exec. log analysis & \citeN{KucherbaevCSCW2016}, \citeN{heymann2011turkalytics}, \citeN{Huynh2013}, \citeN{Jung2014}, \citeN{KhudaBukhsh2014}, \rev{\citeN{Moshfeghi2016}}  \\ \cline{3-4} 
\multicolumn{1}{|l|}{} & \multicolumn{1}{l|}{} & Content analysis &  \citeN{artz2007survey}, \citeN{difallah2013pick}, \citeN{alagarai2014cognitively}, \rev{\citeN{yang2016modeling}} \\ \cline{3-4} 
\multicolumn{1}{|l|}{} & \multicolumn{1}{l|}{} & Transfer learning & \citeN{mo2013cross}, \citeN{fang2014active}, \citeN{zhao2013transfer}  \\ \cline{3-4} 
\multicolumn{1}{|l|}{} & \multicolumn{1}{l|}{} & Collusion detection &   \citeN{KhudaBukhsh2014}, \citeN{marcus2012counting}, \citeN{Allahbakhsh2014SOCA} \\ \hline
\end{tabular}
\end{table}

\begin{table}[h]
\tiny
\setlength{\tabcolsep}{0.4em} 
\renewcommand{\arraystretch}{1.1}
\centering
\caption{Summary of the exemplary usage of assessment methods by the literature referenced in Section \ref{sec:assessment}.
A bullet $\bullet$ in a cell means the method has been used to measure the respective attribute in the quality model.}
\label{tab:assessment}
\begin{tabular}{lll|l|l|l|l|l|l|l|l|l|l|l|l|l|l|l|l|l|l|l|l|l|l|l|l|l|l|l|l|l|l|l|l|l|l|l|l|l|l|}
\cline{4-41}
 &  &  &   \multicolumn{38}{c|}{{\bf Quality Model}} \\ \cline{4-41} 
 &  &  &   \multicolumn{4}{c|}{\multirow{3}{*}{{\bf Data}}} & \multicolumn{13}{c|}{\multirow{1}{*}{{\bf Task}}} & \multicolumn{21}{c|}{\multirow{1}{*}{{\bf People}}}  \\ \cline{8-41}
 &  &  &   \multicolumn{4}{c|}{} & \multicolumn{2}{c|}{\multirow{2}{*}{{\bf \begin{tabular}[c]{@{}c@{}}Task\\desc.\end{tabular}}}} & \multicolumn{3}{c|}{\multirow{2}{*}{{\bf \begin{tabular}[c]{@{}c@{}}User\\interf.\end{tabular}}}} & \multicolumn{2}{c|}{\multirow{2}{*}{{\bf \begin{tabular}[c]{@{}c@{}}In-\\cent.\end{tabular}}}} & \multicolumn{4}{c|}{\multirow{2}{*}{{\bf \begin{tabular}[c]{@{}c@{}}Terms \&\\cond.\end{tabular}}}} & \multicolumn{2}{c|}{\multirow{2}{*}{{\bf \begin{tabular}[c]{@{}c@{}}Perf.\end{tabular}}}} & \multicolumn{4}{c|}{\multirow{2}{*}{{\bf \begin{tabular}[c]{@{}c@{}}Re-\\quester\end{tabular}}}} & \multicolumn{14}{c|}{\multirow{1}{*}{{\bf Worker}}} & \multicolumn{3}{c|}{\multirow{2}{*}{{\bf Group}}} \\ \cline{25-38}
 &  &  &   \multicolumn{4}{c|}{} & \multicolumn{2}{c|}{} & \multicolumn{3}{c|}{} & \multicolumn{2}{c|}{} & \multicolumn{4}{c|}{} & \multicolumn{2}{c|}{} & \multicolumn{4}{l|}{} & \multicolumn{9}{c|}{\bf Profile} & \multicolumn{2}{c|}{\bf Cr.} & \multicolumn{3}{c|}{\bf Exp.} & \multicolumn{3}{c|}{} \\ \cline{4-41} 
 &  &  &   \rot{Accuracy} & \rot{Consistency} & \rot{Timeliness} & \rot{Ad hoc attributes} & \rot{Clarity} & \rot{Complexity} & \rot{Usability} & \rot{Learnability} & \rot{Robustness} & \rot{Extrinsic} & \rot{Intrinsic} & \rot{Privacy} & \rot{IP protection} & \rot{Information security} & \rot{Compliance} & \rot{Cost efficiency} & \rot{Time efficiency} & \rot{Communicativity} & \rot{Generosity} & \rot{Fairness} & \rot{Promptness} & \rot{Age} & \rot{Gender} & \rot{Location}  & \rot{Openness} & \rot{Conscientiousness} & \rot{Extraversion} & \rot{Agreeableness} & \rot{Neuroticism} & \rot{\rev{Motivation}} & \rot{Skills} & \rot{Certificates} & \rot{Badges} & \rot{Reliability} & \rot{Reputation} & \rot{Availability} & \rot{Diversity} & \rot{Non-collusiveness} \\ \hline
\multicolumn{1}{|l|}{\multirow{23}{*}{{\bf \rot{Quality Assessment}}}} & 
\multicolumn{1}{l|}{\multirow{7}{*}{{\bf \rot{Individual}}}} & Rating & $\bullet$ &  &  & $\bullet$ & $\bullet$ & $\bullet$ &  &  &  & $\bullet$ &  &  &  &  &  &  &  & $\bullet$ & $\bullet$ & $\bullet$ & $\bullet$ &  &  &  &  &  &  &  &  &  &  &  &  & $\bullet$ & $\bullet$ &  &  &   \\ \cline{3-41}
\multicolumn{1}{|l|}{} & \multicolumn{1}{l|}{} & Qualification test &  &  &  &  &  &  &  &  &  &  &  &  &  &  &  &  &  &  &  &  &  &  &  &  &  &  &  &  &  &  & $\bullet$ &  &  &  &  &  &  &  \\ \cline{3-41} 
\multicolumn{1}{|l|}{} & \multicolumn{1}{l|}{} & Self-assessment & $\bullet$ &  &  & $\bullet$ &  &  &  &  &  &  &  &  &  &  &  &  &  &  &  &  &  &  &  &  &  &  &  &  &  &  & $\bullet$ &  &  & $\bullet$ &  &  &  &  \\ \cline{3-41}
\multicolumn{1}{|l|}{} & \multicolumn{1}{l|}{} & Personality test &  &  &  &  &  &  &  &  &  &  &  &  &  &  &  &  &  &  &  &  &  &  &  &  & $\bullet$ & $\bullet$ & $\bullet$ & $\bullet$ &$\bullet$  &  &  &  &  &  &  &  &  &   \\ \cline{3-41} 
\multicolumn{1}{|l|}{} & \multicolumn{1}{l|}{} & Referrals &  &  &  &  &  &  &  &  &  &  &  &  &  &  &  &  &  &  &  &  &  &  &  &  &  &  &  &  &  &  & $\bullet$ &  &  & $\bullet$ &  &  &  &  \\ \cline{3-41} 
\multicolumn{1}{|l|}{} & \multicolumn{1}{l|}{} & Expert review & $\bullet$ &  &  & $\bullet$ &  &  &  &  &  &  &  &  &  &  &  &  &  &  &  &  &  &  &  &  &  &  &  &  &  &  &  &  &  & $\bullet$ &  &  &  &   \\ \cline{3-41} 
\multicolumn{1}{|l|}{} & \multicolumn{1}{l|}{} & Usability check &  &  &  &  & $\bullet$ & $\bullet$ & $\bullet$ & $\bullet$ &  &  &  &  &  &  &  &  &  &  &  &  &  &  &  &  &  &  &  &  &  &  &  &  &  &  &  &  &  &  \\ \cline{2-41} 
\multicolumn{1}{|l|}{} & \multicolumn{1}{l|}{\multirow{6}{*}{{\bf \rot{Group}}}} & Voting & $\bullet$ &  &  &  &  &  &  &  &  &  &  &  &  &  &  &  &  &  &  &  &  &  &  &  &  &  &  &  &  &  &  &  &  &  &  &  &  &  \\ \cline{3-41}
\multicolumn{1}{|l|}{} & \multicolumn{1}{l|}{} & Group consensus & $\bullet$ &  &  &  &  &  &  &  & $\bullet$ &  &  &  &  &  &  &  &  &  &  &  &  &  &  &  &  &  &  &  &  &  &  &  &  & $\bullet$ &  &  &  &  \\ \cline{3-41} 
\multicolumn{1}{|l|}{} & \multicolumn{1}{l|}{} & Output agreement & $\bullet$ &  &  &  &  &  &  &  &  &  &  &  &  &  &  &  &  &  &  &  &  &  &  &  &  &  &  &  &  &  &  &  &  & $\bullet$ &  &  &  &  \\ \cline{3-41} 
\multicolumn{1}{|l|}{} & \multicolumn{1}{l|}{} & Peer review & $\bullet$ &  &  &  &  &  &  &  & $\bullet$ &  &  &  &  &  &  &  &  &  &  &  &  &  &  &  &  &  &  &  &  &  &  &  &  & $\bullet$ & $\bullet$ &  &  &   \\ \cline{3-41} 
\multicolumn{1}{|l|}{} & \multicolumn{1}{l|}{} & Feedback aggreg. & $\bullet$ &  &  &  &  &  &  &  &  &  &  &  &  &  &  &  &  &  &  & $\bullet$ &  &  &  &  &  &  &  &  &  &  &  &  &  & $\bullet$ & $\bullet$ &  &  &  \\  \cline{3-41} 
\multicolumn{1}{|l|}{} & \multicolumn{1}{l|}{} & User study &  &  &  &  & $\bullet$ & $\bullet$ & $\bullet$ & $\bullet$ &  & $\bullet$ &  &  &  &  &  &  &  &  &  &  &  & $\bullet$ & $\bullet$ & $\bullet$ &  &  &  &  &  &  &  &  &  &  &  &  &  &  \\ \cline{2-41}
\multicolumn{1}{|l|}{} & \multicolumn{1}{l|}{\multirow{10}{*}{{\bf \rot{Computation-based}}}} & Ground truth & $\bullet$ &  &  &  &  &  &  &  &  &  &  &  &  &  &  &  &  &  &  &  &  &  &  &  &  &  &  &  &  &  &  &  &  & $\bullet$ &  &  &  &  \\ \cline{3-41}
\multicolumn{1}{|l|}{} & \multicolumn{1}{l|}{} & Outlier analysis & $\bullet$ &  & $\bullet$ &  &  &  &  &  &  &  &  &  &  &  &  &  &  &  &  &  &  &  &  &  &  &  &  &  &  &  &  &  &  & $\bullet$ &  &  &  &  \\ \cline{3-41} 
\multicolumn{1}{|l|}{} & \multicolumn{1}{l|}{} & Fingerprinting & $\bullet$ &  &  &  &  &  &  &  &  &  &  &  &  &  &  &  &  &  &  &  &  &  &  &  &  &  &  &  &  &  &  &  &  & $\bullet$ &  &  &  & \\ \cline{3-41} 
\multicolumn{1}{|l|}{} & \multicolumn{1}{l|}{} & Achievements &  &  &  &  &  &  &  &  &  &  &  &  &  &  &  &  &  &  &  &  &  &  &  &  &  &  &  &  &  &  &  &  & $\bullet$ &  &  &  &  &  \\ \cline{3-41} 
\multicolumn{1}{|l|}{} & \multicolumn{1}{l|}{} & Implicit feedback &  &  &  &  &  &  &  &  &  &  &  &  &  &  &  &  &  &  &  &  &  &  &  &  &  &  &  &  &  &  &  &  &  & $\bullet$ & $\bullet$ &  &  &   \\ \cline{3-41} 
\multicolumn{1}{|l|}{} & \multicolumn{1}{l|}{} & Association analysis &  &  &  &  &  &  &  &  &  &  &  &  &  &  &  &  &  &  &  &  &  &  &  &  &  &  &  &  &  &  &  &  &  &  & $\bullet$ &  &  &  \\ \cline{3-41} 
\multicolumn{1}{|l|}{} & \multicolumn{1}{l|}{} & Exec. log analysis & $\bullet$ &  & $\bullet$ &  &  &  &  &  &  &  &  &  &  &  &  &  &  &  &  &  &  & $\bullet$ & $\bullet$ & $\bullet$ &  &  &  &  &  &  &  &  &  & $\bullet$ &  &  &   & $\bullet$\\ \cline{3-41} 
\multicolumn{1}{|l|}{} & \multicolumn{1}{l|}{} & Content analysis & $\bullet$ &  &  &  &  & $\bullet$ & $\bullet$ &  &  & $\bullet$ &  &  &  &  &  &  &  &  &  &  &  &  &  &  &  &  &  &  &  &  &  &  &  &  &  &  &  &  \\ \cline{3-41} 
\multicolumn{1}{|l|}{} & \multicolumn{1}{l|}{} & Transfer learning &  &  &  &  &  &  &  &  &  &  &  &  &  &  &  &  &  &  &  &  &  &  & $\bullet$ &  &  &  &  &  &  &  &  $\bullet$ &  &  & $\bullet$ &  &  &  &  \\ \cline{3-41} 
\multicolumn{1}{|l|}{} & \multicolumn{1}{l|}{} & Collusion detection &  &  &  &  &  &  &  &  &  &  &  &  &  &  &  &  &  &  &  &  &  &  &  &  &  &  &  &  &  &  &  &  &  &  &  &  &  & $\bullet$ \\ \hline
\end{tabular}
\\ Perf. = Performance | Cr. = Credentials | Exp. = Experience
\end{table}

%
%

\begin{table}[t]
\scriptsize
\setlength{\tabcolsep}{0.22em} 
\renewcommand{\arraystretch}{1.1}
\centering
\caption{Representative literature references for the identified assurance actions.}
\label{tab:assureliterature}
\begin{tabular}{llll|p{9.5cm}|}
\cline{5-5}
 &  &  &  & {\bf References} \\ \hline 
\multicolumn{1}{|l|}{\multirow{55}{*}{{\bf \rot{Quality Assurance}}}} & 
\multicolumn{2}{c|}{\multirow{7}{*}{{\bf \rot{\begin{tabular}[c]{@{}c@{}}Improve \\data \\ quality\end{tabular}}}}} & Cleanse data & \citeN{Khazankin2012}, \citeN{DBLP:conf/amcis/SchulzeNS13}, \citeN{bozzon2012answering}, \citeN{Bigham:2010}  \\ \cline{4-5} 
\multicolumn{1}{|l|}{} & \multicolumn{2}{l|}{} & Aggregate outputs & \citeN{crowdwisdombook},  \citeN{aydin2014crowdsourcing}, \rev{\citeN{Ho2016}, \citeN{Gao2016}, \citeN{Ok2016}, \citeN{Chao2012}, \citeN{de2015reliable}, \citeN{ukkonen2015crowdsourced}}  \\ \cline{4-5} 
\multicolumn{1}{|l|}{} & \multicolumn{2}{l|}{} & Filter outputs & \citeN{Dow2012}, \citeN{hansen2013quality}, \citeN{Rao2013}, \citeN{marcus2012counting}, \citeN{Jung2012}, \rev{\citeN{Zhuang2015}, \citeN{Wauthier2011}}  \\ \cline{4-5} 
\multicolumn{1}{|l|}{} & \multicolumn{2}{l|}{} & Iterative improvement & \citeN{little2010turkit}, \citeN{little2010exploring},  \citeN{Kulkarni2012} \\ \cline{2-5}
\multicolumn{1}{|l|}{} & \multicolumn{2}{c|}{\multirow{13}{*}{{\bf \rot{\begin{tabular}[c]{@{}c@{}}Select\\people\end{tabular}}}}} & Filter workers & \citeN{Allahbakhsh2013}, \citeN{zhao2013transfer}, \citeN{kazai2012face}, \citeN{li2014wisdom}, \citeN{trustcol12}, \citeN{hara2013combining}, \rev{\citeN{Abraham2016}, \citeN{Nushi2015}}  \\ \cline{4-5} 
\multicolumn{1}{|l|}{} & \multicolumn{2}{l|}{} & Reject workers & \citeN{difallah2012mechanical}, \citeN{lasecki2014information}, \citeN{difallah2012mechanical}, \citeN{Bozzon2013}, \citeN{marcus2012counting}, \citeN{Allahbakhsh2014SOCA}  \\ \cline{4-5} 
\multicolumn{1}{|l|}{} & \multicolumn{2}{l|}{} & Assign workers & \citeN{kulkarni2012b}, \citeN{Allahbakhsh2013}, \citeN{difallah2013pick}, \citeN{Kulkarni2014}, \citeN{ho2012online}, \rev{\citeN{roy2015task}, \citeN{Kobren2015}}  \\ \cline{4-5} 
\multicolumn{1}{|l|}{} & \multicolumn{2}{l|}{} & Recommend tasks & \citeN{Bernstein201OptRealtime}, \citeN{DBLP:journals/npl/YuenKL15}, \citeN{lin2014signals}  \\ \cline{4-5} 
\multicolumn{1}{|l|}{} & \multicolumn{2}{l|}{} & Promote tasks & \citeN{hu2012deploying}, \citeN{kulkarni2012b}, \citeN{nath2012threats}, \rev{\cite{Oosterman2016}}  \\ \cline{4-5} 
\multicolumn{1}{|l|}{} & \multicolumn{2}{l|}{} & Situated crowdsourcing &  \citeN{hosio2014situated}, \citeN{Heimerl:2012:CEL:2207676.2208619}, \citeN{vaish2014twitch}, \rev{\citeN{Niforatos2016}}  \\ \cline{4-5} 
\multicolumn{1}{|l|}{} & \multicolumn{2}{l|}{} & Recruit teams & \citeN{vukovic2010peoplecloud}, \citeN{schall2012expert}, \citeN{retelny2014expert}, \citeN{li2014wisdom}, \rev{\citeN{Rokicki2015}}  \\ \cline{2-5}
\multicolumn{1}{|l|}{} & \multicolumn{1}{l|}{\multirow{13}{*}{{\bf \rot{\begin{tabular}[c]{@{}c@{}}Incentivize\\people\end{tabular}}}}} & \multicolumn{1}{l|}{\multirow{6}{*}{{\bf \rot{Extr.}}}} & Tailor rewards & \citeN{faradani2011s}, \citeN{DBLP:conf/hcomp/MaoKCHSLS13}, \citeN{scekic2013incentives},  \citeN{Sakurai2013}, \citeN{singer2013pricing}, \citeN{rokicki2014competitive}, \rev{\citeN{Ho2015}, \citeN{radanovic2016learning}, \citeN{Ikeda2016}, \citeN{scekic2013incentives}, \citeN{Kobren2015}} \\ \cline{4-5} 
\multicolumn{1}{|l|}{} & \multicolumn{1}{l|}{} & \multicolumn{1}{l|}{} & Pay bonus & \citeN{scekic2013incentives},  \citeN{Difallah2014}, \citeN{Yin2014}, \citeN{yu2014comparison}, \citeN{Faltings2014} \\ \cline{4-5} 
\multicolumn{1}{|l|}{} & \multicolumn{1}{l|}{} & \multicolumn{1}{l|}{} & Promote workers & \citeN{scekic2013incentives}, \citeN{Dow2012}, \citeN{scekic2013programming}  \\ \cline{3-5}
\multicolumn{1}{|l|}{} & \multicolumn{1}{l|}{} & \multicolumn{1}{l|}{\multirow{7}{*}{{\bf \rot{Intr.}}}} & Share purpose & \citeN{dontcheva2014combining}, \citeN{DBLP:conf/hcomp/MaoKCHSLS13}, \rev{\citeN{Kobayashi2015}, \citeN{Kaufman2016}}  \\ \cline{4-5} 
\multicolumn{1}{|l|}{} & \multicolumn{1}{l|}{} & \multicolumn{1}{l|}{} & Self-monitoring &  \citeN{ipeirotis2014quizz}, \citeN{scekic2013incentives}, \citeN{ipeirotis2014quizz}, \citeN{rokicki2014competitive}, \citeN{dontcheva2014combining}, \citeN{preist2014competing} \\ \cline{4-5} 
\multicolumn{1}{|l|}{} & \multicolumn{1}{l|}{} & \multicolumn{1}{l|}{} & Social transparency & \citeN{huang2013don}, \citeN{huang2013don}, \citeN{viegas2007hidden}, \citeN{yu2014comparison}, \rev{\citeN{Feyisetan2016}}  \\ \cline{4-5} 
\multicolumn{1}{|l|}{} & \multicolumn{1}{l|}{} & \multicolumn{1}{l|}{} & \rev{Gamify task} & \rev{\citeN{Ahn2006}, \citeN{KrauseK15}, \citeN{Law2016}, \citeN{Feyisetan2015}}  \\ \cline{2-5}
\multicolumn{1}{|l|}{} & \multicolumn{2}{c|}{\multirow{4}{*}{{\bf \rot{\begin{tabular}[c]{@{}c@{}}Train \\people\end{tabular}}}}} & Prime workers & \citeN{Morris2012}, \citeN {alagarai2014cognitively}, \citeN{Faltings2014}  \\ \cline{4-5} 
\multicolumn{1}{|l|}{} & \multicolumn{2}{l|}{} & Teach workers &  \citeN{dontcheva2014combining}, \citeN{yu2014comparison}, \rev{\citeN{Doroudi2016}, \citeN{Singla2014}}  \\ \cline{4-5} 
\multicolumn{1}{|l|}{} & \multicolumn{2}{l|}{} & Provide feedback & \citeN{Dow2012}, \citeN{zhu2014reviewing}, \citeN{Kulkarni2012}, \citeN{yu2014comparison}  \\ \cline{4-5} 
\multicolumn{1}{|l|}{} & \multicolumn{2}{l|}{} & Team work & \citeN{kittur2010crowdsourcing}, \citeN{andre2014effects},  \citeN{Dorn2012} \\ \cline{2-5}
\multicolumn{1}{|l|}{} & \multicolumn{2}{c|}{\multirow{9}{*}{{\bf \rot{\begin{tabular}[c]{@{}c@{}}Improve \\task \\design\end{tabular}}}}} & Lower complexity & \citeN{rogstadius2011assessment}, \citeN{anderton2013analysis}, \rev{\citeN{Cheng2015b}}  \\ \cline{4-5} 
\multicolumn{1}{|l|}{} & \multicolumn{2}{l|}{} & Decompose task & \citeN{Kittur2011}, \citeN{Kulkarni2012}  \\ \cline{4-5} 
\multicolumn{1}{|l|}{} & \multicolumn{2}{l|}{} & Separate duties & \citeN{bernstein2010soylent}, \citeN{Kulkarni2012}  \\ \cline{4-5} 
\multicolumn{1}{|l|}{} & \multicolumn{2}{l|}{} & Validate worker inputs & \citeN{bragg2013crowdsourcing}, \citeN{quinn2014asksheet}  \\ \cline{4-5} 
\multicolumn{1}{|l|}{} & \multicolumn{2}{l|}{} & Improve usability &  \citeN{nielsen2002homepage}, \citeN{khanna2010evaluating}, \citeN{Kazai2011}, \citeN{willett2012strategies}, \citeN{alagarai2014cognitively}, \citeN{userstudies} \\ \cline{4-5} 
\multicolumn{1}{|l|}{} & \multicolumn{2}{l|}{} & \rev{Prompt for rationale} &  \rev{\citeN{Mcdonnell16}, \citeN{drapeau2016microtalk} } \\ \cline{4-5} 
\multicolumn{1}{|l|}{} & \multicolumn{2}{l|}{} & \rev{Introduce breaks} &  \rev{\citeN{Dai2015}} \\ \cline{4-5} 
\multicolumn{1}{|l|}{} & \multicolumn{2}{l|}{} & \rev{Embrace error} &  \rev{\citeN{Krishna2016}} \\ \cline{2-5}
\multicolumn{1}{|l|}{} & \multicolumn{2}{c|}{\multirow{9}{*}{{\bf \rot{\begin{tabular}[c]{@{}c@{}}Control \\ execution\end{tabular}}}}} & Reserve workers & \citeN{Bernstein201OptRealtime}, \citeN{lasecki2013real}, \citeN{Bigham:2010} \\ \cline{4-5} 
\multicolumn{1}{|l|}{} & \multicolumn{2}{l|}{} & Flood task list & \citeN{Chilton:2010:TSH:1837885.1837889}, \citeN{Bernstein201OptRealtime}  \\ \cline{4-5} 
\multicolumn{1}{|l|}{} & \multicolumn{2}{l|}{} & Dyn. instantiate tasks & \citeN{KucherbaevCSCW2016}, \citeN{Bozzon2013}, \rev{\citeN{yan2011active}, \citeN{Li2016}, \citeN{Tran-ThanhHRRJ15}, \citeN{chen13a}, \citeN{Karger2011}, \citeN{Karger2014}, \citeN{Bansal2016}}  \\ \cline{4-5} 
\multicolumn{1}{|l|}{} & \multicolumn{2}{l|}{} & Control task order &  \citeN{vesdapunt2014crowdsourcing}, \citeN{marcus2012counting}, \rev{\citeN{Lasecki2015}, \citeN{Newell2016}, \citeN{Eickhoff2013}, \citeN{Difallah2016}} \\ \cline{4-5} 
\multicolumn{1}{|l|}{} & \multicolumn{2}{l|}{} & Inter-task coordination &  \citeN{Bozzon2013}, \citeN{little2010turkit}, \citeN{minder2012crowdlang}, \citeN{KucherbaevIC2015} \\ \hline
\end{tabular}
\end{table}

\begin{table}[t]
\tiny
\setlength{\tabcolsep}{0.37em} 
\renewcommand{\arraystretch}{1.1}
\centering
\caption{Summary of the exemplary usage of assurance actions by the literature referenced in Section \ref{sec:assessment}.
A bullet $\bullet$ in a cell means the action has been used to act on the respective attribute in the quality model.}
\label{tab:assurance}
\begin{tabular}{llll|l|l|l|l|l|l|l|l|l|l|l|l|l|l|l|l|l|l|l|l|l|l|l|l|l|l|l|l|l|l|l|l|l|l|l|l|l|l|}
\cline{5-42}
&  &  &  &   \multicolumn{38}{c|}{{\bf Quality Model}} \\ \cline{5-42} 
&  &  &  &   \multicolumn{4}{c|}{\multirow{3}{*}{{\bf Data}}} & \multicolumn{13}{c|}{\multirow{1}{*}{{\bf Task}}} & \multicolumn{21}{c|}{\multirow{1}{*}{{\bf People}}}  \\ \cline{9-42}
&  &  &  &  \multicolumn{4}{c|}{} & \multicolumn{2}{c|}{\multirow{2}{*}{{\bf \begin{tabular}[c]{@{}c@{}}Task\\desc.\end{tabular}}}} & \multicolumn{3}{c|}{\multirow{2}{*}{{\bf \begin{tabular}[c]{@{}c@{}}User\\interf.\end{tabular}}}} & \multicolumn{2}{c|}{\multirow{2}{*}{{\bf \begin{tabular}[c]{@{}c@{}}In-\\cent.\end{tabular}}}} & \multicolumn{4}{c|}{\multirow{2}{*}{{\bf \begin{tabular}[c]{@{}c@{}}Terms \&\\cond.\end{tabular}}}} & \multicolumn{2}{c|}{\multirow{2}{*}{{\bf \begin{tabular}[c]{@{}c@{}}Perf.\end{tabular}}}} & \multicolumn{4}{c|}{\multirow{2}{*}{{\bf \begin{tabular}[c]{@{}c@{}}Re-\\quester\end{tabular}}}} & \multicolumn{14}{c|}{\multirow{1}{*}{{\bf Worker}}} & \multicolumn{3}{c|}{\multirow{2}{*}{{\bf Group}}} \\ \cline{26-39}
&  &  &  &  \multicolumn{4}{c|}{} & \multicolumn{2}{c|}{} & \multicolumn{3}{c|}{} & \multicolumn{2}{c|}{} & \multicolumn{4}{c|}{} & \multicolumn{2}{c|}{} & \multicolumn{4}{l|}{} & \multicolumn{9}{c|}{\bf Profile} & \multicolumn{2}{c|}{\bf Cr.} & \multicolumn{3}{c|}{\bf Exp.} & \multicolumn{3}{c|}{} \\ \cline{5-42} 
&  &  &  &   \rot{Accuracy} & \rot{Consistency} & \rot{Timeliness} & \rot{Ad hoc attributes} & \rot{Clarity} & \rot{Complexity} & \rot{Usability} & \rot{Learnability} & \rot{Robustness} & \rot{Extrinsic} & \rot{Intrinsic} & \rot{Privacy} & \rot{IP protection} & \rot{Information security} & \rot{Compliance} & \rot{Cost efficiency} & \rot{Time efficiency} & \rot{Communicativity} & \rot{Generosity} & \rot{Fairness} & \rot{Promptness} & \rot{Age} & \rot{Gender} & \rot{Location}  & \rot{Openness} & \rot{Conscientiousness} & \rot{Extraversion} & \rot{Agreeableness} & \rot{Neuroticism} & \rot{\rev{Motivation}} & \rot{Skills} & \rot{Certificates} & \rot{Badges} & \rot{Reliability} & \rot{Reputation} & \rot{Availability} & \rot{Diversity} & \rot{Non-collusiveness} \\ \hline
\multicolumn{1}{|l|}{\multirow{28}{*}{{\bf \rot{Quality Assurance}}}} & 
\multicolumn{2}{c|}{\multirow{4}{*}{{\bf \rot{\begin{tabular}[c]{@{}c@{}}Improve \\data\\ quality\end{tabular}}}}} & Cleanse data & $\bullet$ & $\bullet$ &  & $\bullet$ &  &  &  &  &  &  &  &  &  &  &  &  &  &  &  &  &  &  &  &  &  &  &  &  &  &  &  &  &  &  &  &  &  &   \\ \cline{4-42} 
\multicolumn{1}{|l|}{} & \multicolumn{2}{l|}{} & Aggregate outputs & $\bullet$ &  &  &  &  &  &  &  &  &  &  &  &  &  &  &  &  &  &  &  &  &  &  &  &  &  &  &  &  &  &  &  &  &  &  &  &  &  \\ \cline{4-42} 
\multicolumn{1}{|l|}{} & \multicolumn{2}{l|}{} & Filter outputs & $\bullet$ & $\bullet$ &  & $\bullet$ &  &  &  &  &  &  &  &  &  &  &  &  &  &  &  &  &  &  &  &  &  &  &  &  &  &  &  &  &  &  &  &  &  &  \\ \cline{4-42} 
\multicolumn{1}{|l|}{} & \multicolumn{2}{l|}{} & Iterative improvem. & $\bullet$ & $\bullet$ &  & $\bullet$ &  &  &  &  &  &  &  &  &  &  &  &  &  &  &  &  &  &  &  &  &  &  &  &  &  &  &  &  &  &  &  &  &  &  \\ \cline{2-42} 
\multicolumn{1}{|l|}{} & \multicolumn{2}{c|}{\multirow{7}{*}{{\bf \rot{\begin{tabular}[c]{@{}c@{}}Select\\people\end{tabular}}}}} & Filter workers & $\bullet$ &  &  &  &  &  &  &  &  &  &  &  &  &  &  &  &  &  &  &  &  & $\bullet$ & $\bullet$ & $\bullet$ &  &  &  &  &  &  & $\bullet$ & $\bullet$ & $\bullet$ & $\bullet$ & $\bullet$ &  & $\bullet$ &  \\ \cline{4-42} 
\multicolumn{1}{|l|}{} & \multicolumn{2}{l|}{} & Reject workers & $\bullet$ &  &  & $\bullet$ &  &  &  &  &  &  &  &  &  &  &  &  &  &  &  &  &  &  &  &  &  &  &  &  &  &  &  &  &  & $\bullet$ &  &  &  & $\bullet$ \\ \cline{4-42} 
\multicolumn{1}{|l|}{} & \multicolumn{2}{l|}{} & Assign workers & $\bullet$ &  &  &  &  &  &  &  &  &  &  &  &  &  &  & $\bullet$ & $\bullet$ &  &  &  &  &  &  &  &  &  &  &   &  &  &  &  &  &  &  &  &  &  \\ \cline{4-42} 
\multicolumn{1}{|l|}{} & \multicolumn{2}{l|}{} & Recommend tasks & $\bullet$ &  &  &  &  &  &  &  &  &  &  &  &  &  &  &  &  &  &  &  &  &  &  &  &  &  &  &  &  &  &  &  &  &  &  &  &  &  \\ \cline{4-42} 
\multicolumn{1}{|l|}{} & \multicolumn{2}{l|}{} & Promote tasks &  &  &  &  &  &  &  &  &  &  &  &  &  &  &  & $\bullet$ & $\bullet$ &  &  &  &  &  &  &  &  &  &  &  &  &  & $\bullet$ &  &  &  &  & $\bullet$ & $\bullet$ &  \\ \cline{4-42} 
\multicolumn{1}{|l|}{} & \multicolumn{2}{l|}{} & Situated crowds. &  &  &  &  &  &  &  &  &  &  &  &  &  &  &  &  &  &  &  &  &  &  &  &  &  &  &  &  &  &  &  &  &  &  &  & $\bullet$ &  &  \\ \cline{4-42} 
\multicolumn{1}{|l|}{} & \multicolumn{2}{l|}{} & Recruit teams & $\bullet$ &  &  &  &  &  &  &  &  &  &  &  &  &  &  &  &  &  &  &  &  &  &  &  &  &  &  &   &  &  & $\bullet$ &  &  & $\bullet$ &  &  & $\bullet$ & $\bullet$ \\ \cline{2-42} 

\multicolumn{1}{|l|}{} & \multicolumn{1}{l|}{\multirow{8}{*}{{\bf \rot{\begin{tabular}[c]{@{}c@{}}Incentivize\\people\end{tabular}}}}} & \multicolumn{1}{l|}{\multirow{3}{*}{{\bf \rot{Extr.}}}} & Tailor rewards & $\bullet$ &  & $\bullet$ & $\bullet$ &  &  &  &  &  & $\bullet$ &  &  &  &  &  & $\bullet$ & $\bullet$ &  &  &  &  &  &  &  &  &  &  &  &  &  &  &  &  &  &  & $\bullet$ &  &    \\ \cline{4-42} 
\multicolumn{1}{|l|}{} & \multicolumn{1}{l|}{} & \multicolumn{1}{l|}{} & Pay bonus & $\bullet$ &  & $\bullet$ &  &  &  &  &  &  & $\bullet$ &  &  &  &  &  & $\bullet$ & $\bullet$ &  &  &  &  &  &  &  &  &  &  &  &  &  &  &  &  &  &  & $\bullet$ &  &  \\ \cline{4-42} 
\multicolumn{1}{|l|}{} & \multicolumn{1}{l|}{} & \multicolumn{1}{l|}{} & Promote workers &  &  &  &  &  &  &  &  &  & $\bullet$ & $\bullet$ &  &  &  &  &  &  &  &  &  &  &  &  &  &  &  &  &  &  &  &  &  &  & $\bullet$ &  &  &  &  \\ \cline{3-42} 
\multicolumn{1}{|l|}{} & \multicolumn{1}{l|}{} & \multicolumn{1}{l|}{\multirow{4}{*}{{\bf \rot{Intr.}}}} & Share purpose & $\bullet$ &  &  &  &  &  &  &  &  &  & $\bullet$ &  &  &  &  &  &  &  &  &  &  &  &  &  &  &  &  &  &  &  &  &  &  & $\bullet$ &  &  & $\bullet$ & $\bullet$ \\ \cline{4-42} 
\multicolumn{1}{|l|}{} & \multicolumn{1}{l|}{} & \multicolumn{1}{l|}{} & Self-monitoring & $\bullet$ &  &  &  &  &  &  &  &  &  & $\bullet$ &  &  &  &  &  &  &  &  &  &  &  &  &  &  &  &  &  &  &  &  &  &  & $\bullet$ &  &  &  &  \\ \cline{4-42} 
\multicolumn{1}{|l|}{} & \multicolumn{1}{l|}{} & \multicolumn{1}{l|}{} & Social transparency & $\bullet$ &  &  &  &  &  &  &  &  &  & $\bullet$ &  &  &  &  &  &  &  &  &  &  &  &  &  &  &  &  &  &  &  &  &  &  & $\bullet$  &  & $\bullet$ &  &  \\ \cline{4-42} 
\multicolumn{1}{|l|}{} & \multicolumn{1}{l|}{} & \multicolumn{1}{l|}{} & \rev{Gamify task} & $\bullet$ &  &  &  &  &  &  &  &  &  & $\bullet$ &  &  &  &  &  &  &  &  &  &  &  &  &  &  &  &  &  &  &  &  &  &  &  &  & $\bullet$ &  &  \\ \cline{2-42} 
\multicolumn{1}{|l|}{} & \multicolumn{2}{c|}{\multirow{4}{*}{{\bf \rot{\begin{tabular}[c]{@{}c@{}}Train \\people\end{tabular}}}}} & Prime workers & $\bullet$ &  &  &  &  &  &  &  &  &  &  &  &  &  &  & $\bullet$ &  &  &  &  &  &  &  &  &  &  &  &  &  &  &  &  &  &  &  &  &  &   \\ \cline{4-42} 
\multicolumn{1}{|l|}{} & \multicolumn{2}{l|}{} & Teach workers & $\bullet$ &  &  &  &  &  &  &  &  &  & $\bullet$ &  &  &  &  &  &  &  &  &  &  &  &  &  &  &  &  &   &  &  & $\bullet$ &  &  &  &  &  &  & \\ \cline{4-42} 
\multicolumn{1}{|l|}{} & \multicolumn{2}{l|}{} & Provide feedback & $\bullet$ &  &  &  & $\bullet$ &  &  &  &  &  &  &  &  &  &  &  &  &  &  &  &  &  &  &  &  &  &  &  &  &  & $\bullet$ &  &  &  &  &  &   & \\ \cline{4-42} 
\multicolumn{1}{|l|}{} & \multicolumn{2}{l|}{} & Teamwork & $\bullet$ &  &  &  &  &  &  &  &  &  &  &  &  &  &  &  &  &  &  &  &  &  &  &  &  &  &  &   &  &  &  &  &  &  &  &  &  & $\bullet$ \\ \cline{2-42} 
\multicolumn{1}{|l|}{} & \multicolumn{2}{c|}{\multirow{8}{*}{{\bf \rot{\begin{tabular}[c]{@{}c@{}}Improve \\task \\design\end{tabular}}}}} & Lower complexity & $\bullet$ &  &  &  & $\bullet$  & $\bullet$ &  &  &  &  &  & &  &  &  &  &  &   &  &  &  &  &  &  &  &  &  &  &  &  &  &  &  &  &  &  &  &  \\ \cline{4-42} 
\multicolumn{1}{|l|}{} & \multicolumn{2}{l|}{} & Decompose task & $\bullet$ &  &  &  & $\bullet$ & $\bullet$ &  &  &  &  &  &  &  &  &  &  &  &  &  &  &  &  &  &  &  &  &  &  &  &  &  &  &  &  &  &  &  &  \\ \cline{4-42} 
\multicolumn{1}{|l|}{} & \multicolumn{2}{l|}{} & Separate duties & $\bullet$ &  &  &  &  &  &  &  &  &  &  &  &  &  &  &  &  &  &  &  &  &  &  &  &  &  &  &  &  &  &  &  &  &  &  &  &  &  \\ \cline{4-42} 
\multicolumn{1}{|l|}{} & \multicolumn{2}{l|}{} & Validate inputs & $\bullet$ & $\bullet$ &  & $\bullet$ &  &  &  &  &  &  &  &  &  &  &  &  &  &  &  &  &  &  &  &  &  &  &  &  &  &  &  &  &  &  &  &  &  &  \\ \cline{4-42} 
\multicolumn{1}{|l|}{} & \multicolumn{2}{l|}{} & Improve usability & $\bullet$ &  & $\bullet$ &  &  &  & $\bullet$ & $\bullet$ &  &  &  &  &  &  &  &  &  &  &  &  &  &  &  &  &  &  &  &   &  &  &  &  &  &  &  &  &  &  \\ \cline{4-42} 
\multicolumn{1}{|l|}{} & \multicolumn{2}{l|}{} & \rev{Prompt for rationale} & $\bullet$ &  &  &  &  &  &  &  &  &  &  &  &  &  &  &  &  &  &  &  &  &  &  &  &  &  &  &  &  &  &  &  &  & $\bullet$ &  &  &  &  \\ \cline{4-42} 
\multicolumn{1}{|l|}{} & \multicolumn{2}{l|}{} & \rev{Introduce breaks} & $\bullet$ &  &  &  &  &  &  &  &  &  &  &  &  &  &  &  &  &  &  &  &  &  &  &  &  &  &  &  &  &  &  &  &  &  &  & $\bullet$ &  &  \\ \cline{4-42} 
\multicolumn{1}{|l|}{} & \multicolumn{2}{l|}{} & \rev{Embrace error} &  &  & $\bullet$ &  &  &  &  &  &  &  &  &  &  &  &  &  & $\bullet$ &  &  &  &  &  &  &  &  &  &  &  &  &  &  &  &  &  &  &  &  &  \\ \cline{2-42} 
\multicolumn{1}{|l|}{} & \multicolumn{2}{c|}{\multirow{5}{*}{{\bf \rot{\begin{tabular}[c]{@{}c@{}}Control \\ execution\end{tabular}}}}} & Reserve workers &  &  & $\bullet$ &  &  &  &  &  &  &  &  &  &  &  &  &  &  &  &  &  &  &  &  &  &  &  &  &  &  &  &  &  &  &  &  & $\bullet$ &  &  \\ \cline{4-42} 
\multicolumn{1}{|l|}{} & \multicolumn{2}{l|}{} & Flood task list &  &  & $\bullet$ &  &  &  &  &  &  &  &  &  &  &  &  &  &  &  &  &  &  &  &  &  &  &  &  &  &  &  &  &  &  &  &  & $\bullet$ &  &  \\ \cline{4-42} 
\multicolumn{1}{|l|}{} & \multicolumn{2}{l|}{} & Dyn. inst. tasks & $\bullet$  &  & $\bullet$ &  &  &  &  &  &  &  &  &  &  &  &  & $\bullet$ & $\bullet$ &  &  &  &  &  &  &  &  &  &  &  &  &  &  &  &  &  &  &  &  &  \\ \cline{4-42} 
\multicolumn{1}{|l|}{} & \multicolumn{2}{l|}{} & Control task order & $\bullet$ &  &  &  &  &  &  &  &  &  &  &  &  &  &  & $\bullet$ & $\bullet$ &  &  &  &  &  &  &  &  &  &  &  &  &  &  &  &  &  &  &  &  &  \\ \cline{4-42} 
\multicolumn{1}{|l|}{} & \multicolumn{2}{l|}{} & Inter-task coord. &  &  &  &  &  & $\bullet$ &  &  &  &  &  &  &  &  &  & $\bullet$ & $\bullet$ &  &  &  &  &  &  &  &  &  &  &  &  &  &  &  &  &  &  &  &  &   \\ \hline
\end{tabular}
\end{table}